\documentclass[%
 reprint,
 amsmath,amssymb,
 aps,
]{revtex4-1}

\usepackage{graphicx}
\usepackage{dcolumn}
\usepackage{bm}
\usepackage{xcolor}

\begin{document}

\preprint{APS/123-QED}

\title{Speckled cross-spectral densities and their associated correlation singularities for a modern source of partially coherent x rays}


\author{David M. Paganin}
\email{david.paganin@monash.edu}
\affiliation{School of Physics and Astronomy, Monash University, Victoria 3800, Australia}

\author{Manuel S\'{a}nchez del R\'{i}o}
\email{srio@esrf.eu}
\affiliation{European Synchrotron Radiation Facility, 38043 Grenoble, France}

\date{\today}

\begin{abstract}
We consider a realistic model for calculating the cross-spectral density of partially coherent beams from an x-ray undulator in a modern storage ring.  This two-point coherence function is seen to have a speckled structure associated with the presence of x-ray coherence vortices and domain walls.  Such cross-spectral density speckle is associated with a network of spatial pairs of points for which there is zero correlation.  X-ray coherence vortices and domain walls are seen to emerge naturally as the number of coherent modes required increases.  An understanding of the existence and nature of such correlation singularities enhances our ability to exploit partially coherent x-ray radiation from new or upgraded synchrotron sources, for both imaging and diffraction applications.
\end{abstract}

\maketitle                        


\section{Introduction}

Synchrotron radiation (SR) has witnessed enormous growth in recent decades, largely due to its applicability to multidisciplinary applied science. In particular, many experimental techniques have recently (or relatively recently) been developed to exploit the coherence of SR, such as x-ray photon correlation spectroscopy \cite{XPCS}, coherent diffraction imaging \cite{CDI}, propagation-based phase contrast imaging \cite{PBPCI} and ptychography \cite{Ptychography}. 

Many storage-ring based x-ray synchrotron facilities are building or planning upgrades to increase brilliance and coherent flux by one to three orders of magnitude.  The first upgrade of a large facility will be the EBS (Extremely Brilliant Source) \cite{orangebook} at the European Synchrotron Radiation Facility (ESRF), aiming to build a storage ring of 150 pm emittance to significantly boost the associated x-ray coherence.

Accurate calculation and quantitative evaluation of the parameters related to x-ray coherence, in such new storage rings, is of paramount importance for designing, building and exploiting the new beamlines. In this context an algorithm to calculate the cross-spectral density (CSD) of radiation emitted by modern x-ray undulators has been developed \cite{glass}.  The CSD quantifies two-point correlation properties of partially coherent statistically stationary fields \cite{Wolf1982,mandel_wolf}.  For Gaussian statistics, the CSD completely characterizes the properties of a beam since the Gaussian moment theorem implies all higher-order correlation functions either (i) vanish or (ii) are expressible in terms of the two-point correlation.  Two-point correlation functions are therefore a key input to model x-ray experiments in which x-ray CSDs are streamed through subsequent optical elements and samples, and finally through to the detected spectral density.  

We consider the role played by the CSD phase \cite{Schouten2003,GburVisser2003,Bogatyryova2003}. This governs the position of Young-type interference fringes formed when the disturbance from two different spatial points is combined at a given angular frequency \cite{mandel_wolf}.  It thereby influences detected spectral densities in both imaging and non-imaging contexts.  

As discussed later, this CSD phase can and typically will possess a network of both (i) domain walls across which the CSD phase is discontinuous, and (ii) branch lines around which it has a non-zero winding number.  At either of these CSD phase singularities, the CSD vanishes \cite{TopologicalReactionsCohVortices,Marasinghe2010}.  Thus, even for highly coherent sources such as that considered in the present paper, a complicated network of infinitely many pairs of points will typically exist, for which the x-ray disturbance is totally uncorrelated. 

While such ``correlation singularities'' have received attention in a visible-light setting \cite{Schouten2003,GburVisser2003,Bogatyryova2003,FischerVisser2004,Palacios2004,GburVisser2006,Wang2006,TopologicalReactionsCohVortices,GburVisser2010,Marasinghe2010,Marasinghe2011}, relatively little work exists in an x-ray context.  An exception is a model for a partially coherent x-ray source, which contains embedded correlation singularities in its CSD phase  \cite{PellicciaPaganin2012}. Note also that the Schell model of partially coherent scalar fields \cite{mandel_wolf}, which has been applied to x rays \cite{Coisson1997,Vartanyants2010}, can contain embedded CSD correlation singularities when generalized e.g.~to Laguerre--Gauss Schell beams \cite{Palacios2004,Rodrigo2015}.  Gauss--Schell beams do not contain correlation singularities, however when such beams pass through samples such singularities may develop. 

One key aspect of correlation singularities must be emphasized from the outset.  While it is very well known that complex scalar wavefields vanish at phase singularities such as coherent phase vortices \cite{Dirac1931} and phase domain walls \cite{paganin_book}, the {\em correlation} singularity---which is associated with a {\em vanishing correlation between a  pair of points}, is not in general associated with a field zero \cite{GburVisser2003, GburVisserWolf2004,GburVisser2006,TopologicalReactionsCohVortices,GburVisser2010}.  Gbur and Visser note in this context that ``the intensity of the field at such a pair of points is not required to vanish, and in general will not ... a coherence vortex cannot be associated with any single point of a wavefield, but only pairs of points; it might be said that it is a `virtual' feature of the wavefield'' \cite{GburVisser2003}.   It is for precisely these reasons that Gbur, Visser and Wolf speak of correlation singularities as ``hidden'' singularities \cite{GburVisserWolf2004} whose presence is not heralded by wavefield zeroes, in stark contrast to their coherent counterparts.  The relation between coherent phase singularities (wavefield phase vortices and wavefield domain walls) and correlation singularities (coherence domain walls and coherence vortices) is rather indirect \cite{GburVisser2006,TopologicalReactionsCohVortices,GburVisser2010}, with this relation being a topic of ongoing research \cite{PangGburVisser2015}.

In the visible-light regime, there are several   experimental studies on correlation singularities. These experiments typically deliberately engineer a vortical correlation singularity, e.g.~via a Laguerre--Gauss ${\textrm{LG}}_{0 \pm 1}$ vortex mode \cite{Bogatyryova2003}, spiral phase plate \cite{Palacios2004},  spiral zone plate \cite{Wang2006} or multiple-wave interference \cite{Ambrosini2005} (see also Basano et al.~\cite{Basano2005}, for an acoustic analog).  This gives coherent vortices in the limit of strict monochromaticity, which evolve into correlation singularities when the radiation is partially coherent.  This approach may be contrasted with the view, aligned with the optical concept of ``natural focusing and fine structure'' \cite{Nye1999}, whereby fine wavefield structures such as coherent vortices---together with their partially coherent generalizations, the correlation singularities---may be naturally or spontaneously formed, rather than needing to be deliberately engineered.  Examples of spontaneously nucleated correlation singularities include those associated with the two-pinhole interferometer \cite{Schouten2003}, focused partially coherent light \cite{FischerVisser2004} and a certain simple model for paraxial partially coherent fields in the absence of any sources or optical elements \cite{PellicciaPaganin2012}.

In the present study we find spontaneously-nucleated correlation singularities in the CSD, together with an associated speckled CSD structure, to be implied by a realistic modern x-ray undulator model. This is of practical importance since such correlation singularities can have subtle effects on both imaging and diffraction data. For example, we predict a suppression in the visibility of both the near-field and far-field interference fringes that we would otherwise expect when radiation from two scattering centers separated by less than a coherence width is overlapped, if these two centers coincide with a pair of spatial positions associated with a correlation singularity.  This will be of particular influence in the inside-source method for x-ray Fourier holography \cite{FaigelTegze1999}, but will also apply e.g.~to pairs of points within an extended scattering volume under the first Born approximation, as well as scattering from multi-layer mirrors. As another example of the practical importance of x-ray correlation singularities, the transverse location of interference fringes in Young-type interference experiments---such as may be used to measure the coherence properties of an x-ray beam \cite{Ditmire1996,Takayama1998,Leitenberger1,Leitenberger2,Leitenberger3}---may cycle through all possible locations for very similar sets of two pinholes, and therefore jump unexpectedly for particular sets of pinhole positions, if the pair of pinholes lies close to the sub-class of correlation singularities known as a coherence vortex.  It is also of practical importance that correlation singularities can be formed spontaneously for focused fields \cite{FischerVisser2004} and are therefore to be expected in focused x-ray nano-probes, particularly when aberrations are present.

The broader conclusions and formalism of the present work extend beyond x-ray and visible light optics, to any partially coherent paraxial complex scalar fields obeying the Helmholtz equation in the monochromatic (mono energetic) limit.  As such, all x-ray applications considered above have directly analogous applications in visible-light optics, electron optics and neutron optics.

This paper is structured as follows. Section 2 reviews relevant background regarding the cross-spectral density, including its coherent-mode representation and subsequent numerical evaluation for partially coherent sources.  This section also briefly reviews relevant background on coherence vortices and coherence domain walls, these being the two key correlation-function phase singularities for complex scalar fields.  All of this is formulated using the space--frequency description of partially coherent complex scalar electromagnetic fields.  Section~3 presents a numerical study of the cross-spectral density associated with an x-ray undulator, calculated at the source position.  Coherence vortices and domain walls, together with an associated speckled structure in the CSD, are seen to arise. Section 4 discusses propagation of the CSD calculated at the source, to different distances in the near field, intermediate field and far field. Section 5 discusses broader implications of this work, and outlines avenues for future research.  We conclude with Sec.~6.

\section{Background}

This section is divided into two parts.  We begin by briefly reviewing some relevant background regarding the space--frequency description of partially coherent scalar electromagnetic fields.  We then review some basic results regarding coherence vortices and domain walls in the cross-spectral density of partially coherent fields.  

\subsection{Cross-spectral density for x-ray synchrotron radiation}

The SR emitted by present storage rings is partially coherent, due to superposed single-electron emission. A single electron emits a spatially coherent wave-front that can be calculated using classical electrodynamics \cite{jackson}. However, emission of individual electrons is incoherent among distinct electrons, since the bunch length greatly exceeds the radiation wavelength.  This latter point is always the case for storage-ring x-ray sources, but not for x-ray Free Electron Lasers. The fact that storage ring emittance is low implies the bunch transverse sizes to be small. An observer placed at a sufficiently long distance from the source (typical beamline lengths are 30--200 m) would observe  radiation with a relatively high degree of  spatial coherence. Furthermore, if required, a high degree of temporal coherence can be obtained via a monochromator.  Whether or not the x-ray beam is energy filtered, its partial coherence is due to the statistical distribution of the electrons in the storage ring. 

To completely describe second-order partial coherence properties, the CSD \cite{Wolf1982,mandel_wolf} may be used: 
\begin{eqnarray}
\nonumber W(x_1,y_1,z_1,x_2,y_2,z_2,\omega) \quad\quad\quad\quad\quad\quad \\ = \langle E^{*}(x_1,y_1,z_1,\omega) E(x_2,y_2,z_2,\omega) \rangle.\label{eq:CSD}
\end{eqnarray}
Here, $E$ is the complex (scalar) electric field at two spatial points $\vec{r}_1=(x_1,y_1,z_1)$ and $\vec{r}_2=(x_2,y_2,z_2)$, $\hbar\omega$ is the photon energy and $\hbar$ is the reduced Planck constant $h/(2\pi)$.  We follow the usual convention that $z$ denotes the optic axis with respect to which the electromagnetic disturbance is paraxial. The average, denoted by angular brackets above, is over a Gibbs-like statistical ensemble of strictly monochromatic fields, all of which have the same angular frequency $\omega$. We have implicitly assumed wide-sense statistical stationarity, which is satisfied by emission from storage rings \cite{geloni}. Also, the two observation points are usually in a plane perpendicular to the beam at a distance $z=z_1=z_2$ from the source. Therefore, for practical purposes the CSD is a four-dimensional function for a given $z$ and $\omega$: 
\begin{equation}
W(x_1,y_1,x_2,y_2,z,\omega) = 
\langle E^{*}(x_1,y_1,z,\omega) E(x_2,y_2,z,\omega)\rangle.
\end{equation}

Calculation of the CSD can be performed knowing the distribution of the electrons and the characteristics of the synchrotron-radiation emission. The ``convolution theorem'' of Kim \cite{kim} gives a practical procedure for calculating Wigner functions \cite{Wigner1932} and also the CSD. Note that one-to-one mappings exist that transform the Wigner-function representation of the second-order coherence properties of a partially coherent field, to the CSD representation, and vice versa \cite{AlonsoWignerFunctionReview}.  However  numerical evaluation and storage for either representation is very computationally expensive, as we need to sample such two-point correlation functions with high resolution (about $10^3$ samples per dimension, leading typically to on the order of $10^{12}$ complex numbers, requiring gigabytes to terabytes of computer memory storage). Moreover, propagation in vacuum of the CSD from a plane with fixed $z_1=z_2= z'$ to another plane with $z_1=z_2=z''>z'$ must be done using 4D integrals with the corresponding Green functions \cite{mandel_wolf}, which is certainly beyond the possibilities of present computers in scenarios with realistic levels of complexity. 

A significantly more efficient means to store the CSD uses the property that it can be represented in terms of eigenvalues $\lambda_j(\omega)$ and coherent modes $\psi_j(\vec{r}, \omega)$, via the coherent-mode expansion
\cite{Wolf1982,mandel_wolf}:
\begin{equation}
\begin{aligned}
W(\vec{r_1}, \vec{r_2}, \omega)
=
\sum_{j=0}^{\infty}
\lambda_j(\omega) \,
\psi_j^*(\vec{r_1},\omega)
\psi_j(\vec{r_2}, \omega).
\end{aligned}\label{eq:CoherentModeExpansion}
\end{equation}

It can be shown that a fully coherent beam has a single coherent mode (index $j$ of zero). For a partially coherent beam with high coherent fraction the first modes contain a large fraction of the total spectral density, so a truncated series is a good representation of the CSD. The main advantage of this expansion is that the 4D CSD can be computed as a (truncated) sum of 2D modes, therefore making storage possible. Also, the propagated CSD can be computed from the sum of the propagated modes, each of which propagate in the same manner as a strictly coherent field.  Therefore we can propagate the CSD by performing 2D integrals (for each mode in the CSD) instead of 4D integrals (directly for CSD itself).  

The eigenvalues and associated coherent modes may be obtained as solutions of the following homogeneous Fredholm integral equation of second kind:
\begin{equation}
\begin{aligned}
\label{fredholm_equation}
\int \!\!\! \int \!\!\! \int d^3\vec{r_1}
W(\vec{r_1},\vec{r_2}, \omega){\psi}_m(\vec{r_1},
\omega)=\lambda_m(\omega)\psi_m(\vec{r_2}, \omega).
\end{aligned}
\end{equation}
These solutions may then be assembled into the coherent-mode representation of the CSD \cite{Wolf1982}.  This can be solved numerically, e.g.~using the computer package ``COherent Modes for SYnchrotron Light'' (COMSYL) \cite{glass}.

\subsection{Coherence vortices in the cross-spectral density}

Assume that the medium, through which a given statistically-stationary partially coherent forward-propagating complex scalar electromagnetic field propagates, contains no discontinuities for any of the points $(x_1,y_1,z)$ and $(x_2,y_2,z)$ at which the cross-spectral density $W(x_1,y_1,x_2,y_2,z,\omega)$ is to be computed. This implies $W$ to be a continuous and single-valued complex function of all its arguments.  Using reasoning closely related to that of Dirac in a different context \cite{Dirac1931}, important general CSD properties may be derived from the assumption of single-valuedness and continuity, without needing to make any specific reference to the underpinning  equations that govern the CSD \cite{Marasinghe2010}. 

\begin{figure}
\includegraphics[width=8.5cm]{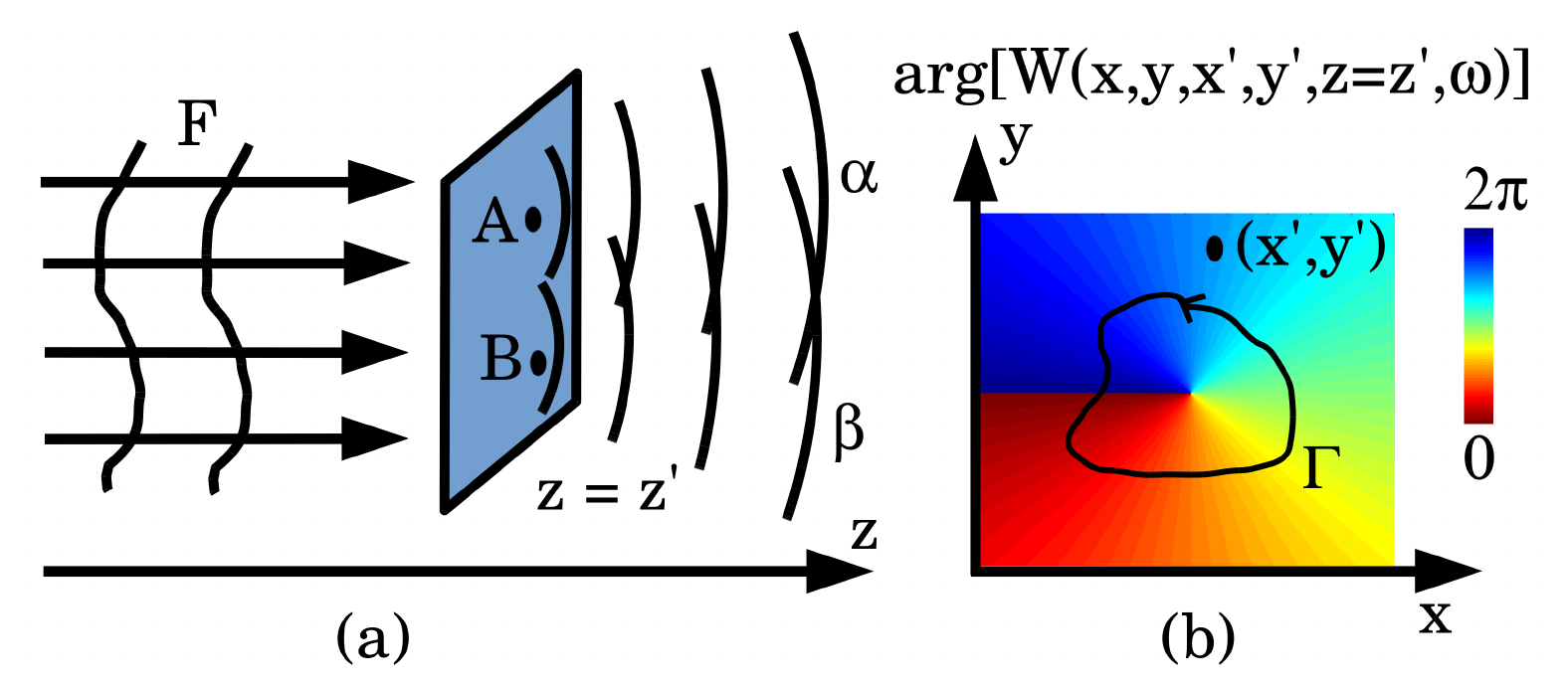}
\caption{(a) The cross-spectral density $W(x,y,x',y',z=z',\omega)$ associated with a statistically stationary field $F$ quantifies the degree of correlation of the disturbance at points $(x,y,z)\equiv A$ and $(x',y',z'=z)\equiv B$. (b) For fixed $(x',y')$, $\omega$ and $z=z'$, the phase (arg) of $W(x,y,x',y',z=z',\omega)$ has a coherence vortex as a function of $x$ and $y$, at a point enclosed within the contour $\Gamma$, about which the phase winds by an integer multiple $m$ of $2\pi$ radians. Here, $m=1$.  See Eq.~\ref{phase_of_W_winding}.}
\label{loss_of_fringe_visibility}
\end{figure}

Let the CSD phase be denoted by:
\begin{equation}
\begin{aligned}
\label{phase_of_W}
\Phi(x_1,y_1,x_2,y_2,z,\omega)=\arg[W(x_1,y_1,x_2,y_2,z,\omega)]
\end{aligned}
\end{equation}
This quantity governs the position of Young-type interference fringes which would be formed, if one were to combine the disturbances $\alpha$ and $\beta$ scattered from points $(x_1,y_1,z)\equiv A$ and $(x_2,y_2,z)\equiv B$, respectively, at energy $\hbar\omega$ \cite{mandel_wolf}.  See Fig.~\ref{loss_of_fringe_visibility}a. The  CSD phase is complemented by the magnitude of the CSD, which governs the visibility of the previously-mentioned Young-type fringes.

While $W\in\mathbb{C}$ and $|W|\in\mathbb{R}$ are both single-valued, $\Phi$ will in general be multi-valued. Indeed, since the phase of a complex number is only defined modulo $2\pi$, it may wind by an integer multiple $m$ of $2\pi$ (see Fig.~\ref{loss_of_fringe_visibility}b) \cite{GburVisser2003,Marasinghe2011}: 
\begin{equation}
\begin{aligned}
\label{phase_of_W_winding}
\oint_{\Gamma} d\Phi(x_1,y_1,x_2,y_2,z,\omega)=2\pi m.
\end{aligned}
\end{equation}
Here, $\Gamma$ is any simple smooth closed clockwise-traversed one-dimensional curve embedded in the six-dimensional space with coordinates $(x_1,y_1,x_2,y_2,z,\omega)$, and $d\Phi$ is the increment in $\Phi$ corresponding to an infinitesimal line segment of $\Gamma$.  Admissible curves $\Gamma$ are those for which $W$ is non-zero at every point on $\Gamma$, ensuring $\Phi$ and hence $d\Phi$ to be well defined at each point on $\Gamma$.  We consider $\omega$ to have a fixed arbitrary value throughout the paper. Non-zero $m$ indicates non-trivial topology in the phase of $W$, with the corresponding phase map in Fig.~\ref{loss_of_fringe_visibility}b being one of infinitely many Riemann sheets describing a screw dislocation threaded by a coherence-vortex core. 

Non-zero $m$ indicates a coherence vortex \cite{GburVisser2003} to be present, with topological charge given by the integer $m$.  Such structures are a partially-coherent analog of phase vortices that may form in the phase of coherent optical fields \cite{NyeBerry1974,Nye1999,SoskinVasnetsov2001,DennisProgOpt2009}, including x-ray fields \cite{XRayVortex1,XRayVortex2,XRayVortex3,XRayVortex4,paganin_book,XRayVortex5,XRayVortex6,XRayVortex7}, together with electron fields \cite{MessiahBook,AllenOxleyPaganin2001, Petersen2013, Bliokh2017} and matter-wave fields \cite{PitaevskiiStringariBook} etc.  Two key consequences of non-vanishing $m$ are outlined below.  These arguments are topological in nature, regarding generic structures that are stable with respect to continuous deformation of the underlying fields. 

The existence of any {\em one} circuit $\Gamma$ for which $m$ is non-zero, implies the presence of a (nodal) manifold of points in $(x_1,y_1,x_2,y_2,z,\omega)$-space, at each of which $W$ vanishes \cite{Marasinghe2010,Marasinghe2011}.  This is remarkable, since vanishing $W$ corresponds to the disturbance at two points being totally incoherent (i.e.~totally uncorrelated), even though the field may have a very high state of coherence. Such points are termed correlation singularities, nodal points, nodal lines, or a nodal manifold.  

If $\Gamma$ lies in a particular two-dimensional hyper-plane $\Pi$ with coordinates $(\xi,\eta)$ within $(x_1,y_1,x_2,y_2,z,\omega)$-space, the coherence vortex will typically occur at a (zero-dimensional) nodal point $(\xi_0,\eta_0)$ in the said hyper-plane, serving as a branch point for $\Phi$, about which $\Phi$ winds by $2 \pi m$ radians (cf.~the $m=1$ case in Fig.~\ref{loss_of_fringe_visibility}b).  The set of nodal points becomes a (one-dimensional) nodal line, or a connected set of nodal lines which either form closed loops or extend to the boundary of the considered region, when we consider the one-dimensional loop $\Gamma$ to be embedded within a three-dimensional hyper-plane in  $(x_1,y_1,x_2,y_2,z,\omega)$-space.  This set of nodal lines (any point on which is a singular point for $\Phi$) may form a tree-like structure, e.g.~if an $m=2$ coherence vortex decays to a pair of $m=1$ coherence vortices \cite{TopologicalReactionsCohVortices,GburSPIE} (cf.~an analogous phenomenon for vortices in the phase of coherent fields \cite{Freund1999}).  Knotted and braided coherence-vortex nodal lines in $W$ are also topologically possible, albeit exotic.  Permissible nodes in tree-like nodal-line structures in $W$ are governed by the law of conservation of topological charge.  This set of nodal lines becomes a two-dimensional network of nodal sheets (zero sheets) in any four-dimensional hyper-plane within the $(x_1,y_1,x_2,y_2,z,\omega)$-space, and a three-dimensional manifold of nodal points in any five-dimensional hyper-plane subset of $(x_1,y_1,x_2,y_2,z,\omega)$-space.  Finally, in the full six-dimensional $(x_1,y_1,x_2,y_2,z,\omega)$-space, the set of nodal points of $W$ will form a four-dimensional network of point pairs $\Upsilon$, at each of which $W$ vanishes \cite{Marasinghe2010}. 

\begin{figure}
\includegraphics[width=8.5cm]{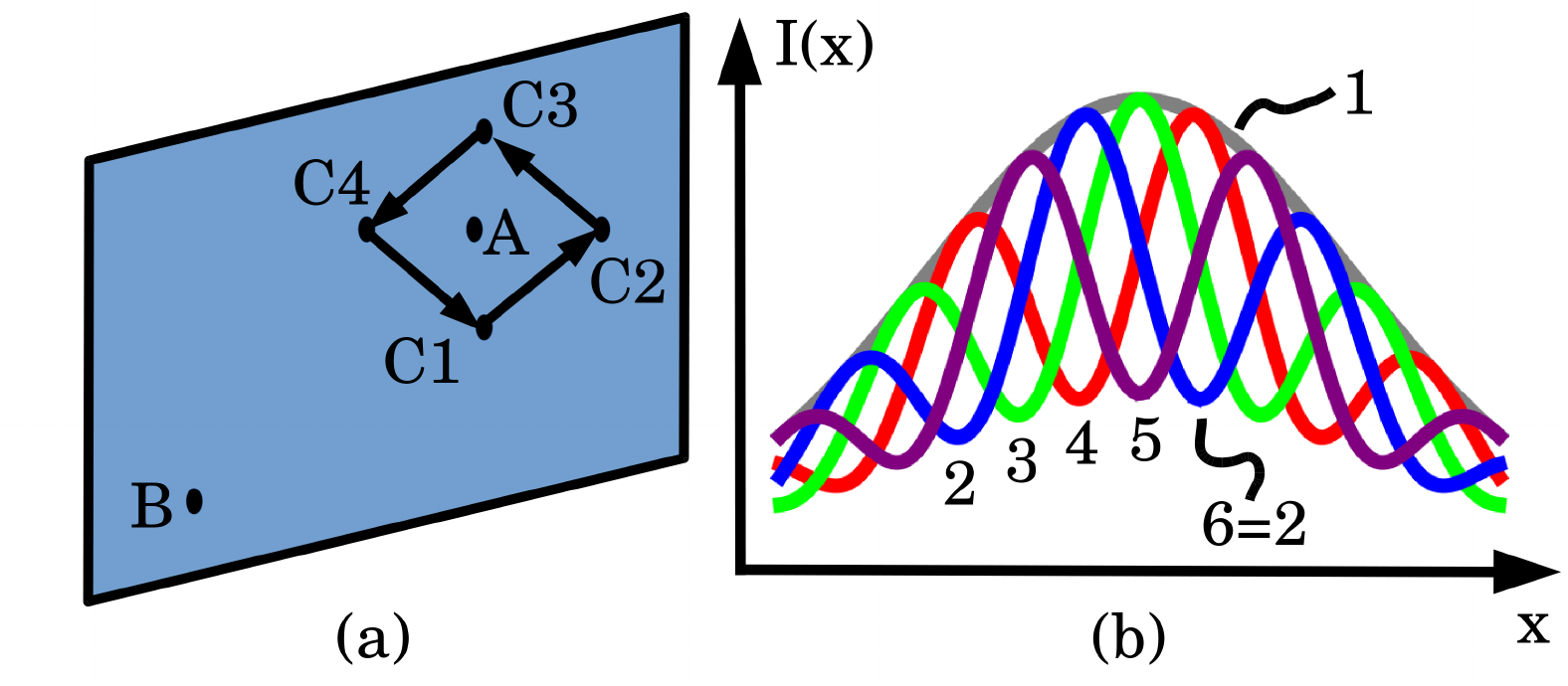}
\caption{Ratcheting of Young interferograms associated with $m=1$ coherence vortex corresponding to the pair of points $A,B$.  (a) A series of Young interferometers is constructed.  In all setups, radiation illuminates a screen in which the first of two pinholes is always at $B$.  The second pinhole is placed at $A$, before being moved through the cycle of locations $C1 \rightarrow C2 \rightarrow C3 \rightarrow C4 \rightarrow C1$.  (b) The resulting interferograms, over some plane downstream of the point scatterers, are shown in curves 1 (pinholes $A$ and $B$), 2 ($C1$ and $B$), 3 ($C2$ and $B$), 4 ($C3$ and $B$), 5 ($C4$ and $B$) and 6 ($C1$ and $B$) respectively. Note that curves 2 and 6 are identical.  Also, $I(x)$ denotes the spectral density of the interferogram as a function of the transverse coordinate $x$, perpendicular to the optic axis.}
\label{Young_fringe_anholonomy}
\end{figure}

These nodal points in $W$ exhibit ``complete destructive interference of coherence''.  More precisely, any point $(x_1',y_1',x_2',y_2',z',\omega)\in\Upsilon$ will correspond to a pair of spatial points $(x_1',y_1',z')\equiv A$ and $(x_2',y_2',z')\equiv B$ for which the partially-coherent disturbance is completely uncorrelated at energy $\hbar\omega$ \cite{Schouten2003,GburVisser2003,Bogatyryova2003}.  If e.g.~a point scatterer were to be placed at $A$, with another point scatterer at $B$, and the radiation scattered from both points allowed to overlap, no interference fringes would be observed at energy $\hbar\omega$.  See the gray curve (curve 1) in Fig.~\ref{Young_fringe_anholonomy}.  The nodal  manifold  $\Upsilon$, which will typically permeate much of the $(x_1,y_1,x_2,y_2,z,\omega)$-space associated with cross-spectral densities calculated for non-trivial systems \cite{TopologicalReactionsCohVortices,GburVisser2003}, is ``a web of incoherence'' spun through a partially-coherent field's cross-spectral density. Note that such a web does not exist in physical space, since correlation singularities do not occur at particular points in space, but rather for pairs of spatial points \cite{GburVisser2003}.   

A second key consequence of the nodal manifold $\Upsilon$ is again related to the visibility of interference fringes associated with pairs of scattering points.  As previously stated, when pinholes are placed at locations $A$ and $B$ in Fig.~\ref{loss_of_fringe_visibility} that correspond to a coherence vortex (see also Fig.~\ref{Young_fringe_anholonomy}a), there is zero fringe visibility in the associated Young-type pattern filtered to energy $\hbar \omega$: see, once more, curve 1 in Fig.~\ref{Young_fringe_anholonomy}b.  However, the non-zero fringe visibility is regained if we move ``off the coherence vortex'' by shifting the first pinhole from location $A$ to location $C1$: see curve 2 in Fig.~\ref{Young_fringe_anholonomy}b.  The vortical nature of the coherence vortex is then evident if we perform a sequence of Young-type interference experiments, where the pinhole at $B$ is kept fixed, while the pinhole at $C1$ is moved through the cycle $C1$ (curve 2)$\rightarrow C2$ (curve 3)$\rightarrow C3$ (curve 4)$\rightarrow C4$ (curve 5)$\rightarrow C1$ (curve 6 = curve 2). If we trace the evolution of the intensity maxima associated with the resulting sequence of interferograms in curves 2 to 6, the physical meaning of $m$ becomes clear: during the cycle, if $m=1$ then the maxima of the interferograms  ``ratchet'' to the right by one fringe during the cycle. If $m=-1$ they would instead ratchet to the left by one fringe during the cycle.  For general $m$, fringes  ratchet to the right (left) by $|m|$ fringes, if $m$ is positive (negative) \cite{Marasinghe2010}.

In addition to the coherence vortex, a second type of CSD phase singularity exists: domain walls.  These were the first-discovered form of CSD  singularity \cite{Schouten2003}.  Such defects are stable for real correlation functions, but not for complex correlation functions such as the CSD.  Nevertheless CSD phase domain walls may exist, especially if a small number of coherent modes is present, with $\pi$ phase shift at points where the CSD changes sign.  As with the coherence vortex, $|W|=0$ at CSD domain walls.  The dimension of a domain-wall CSD singularity network in $(x_1,y_1,x_2,y_2,z,\omega)$-space, if it exists, is one higher than for a corresponding coherence-vortex network.  

\section{Simulations of unpropagated x-ray undulator beams}

\begin{figure}
\includegraphics[width=8cm]{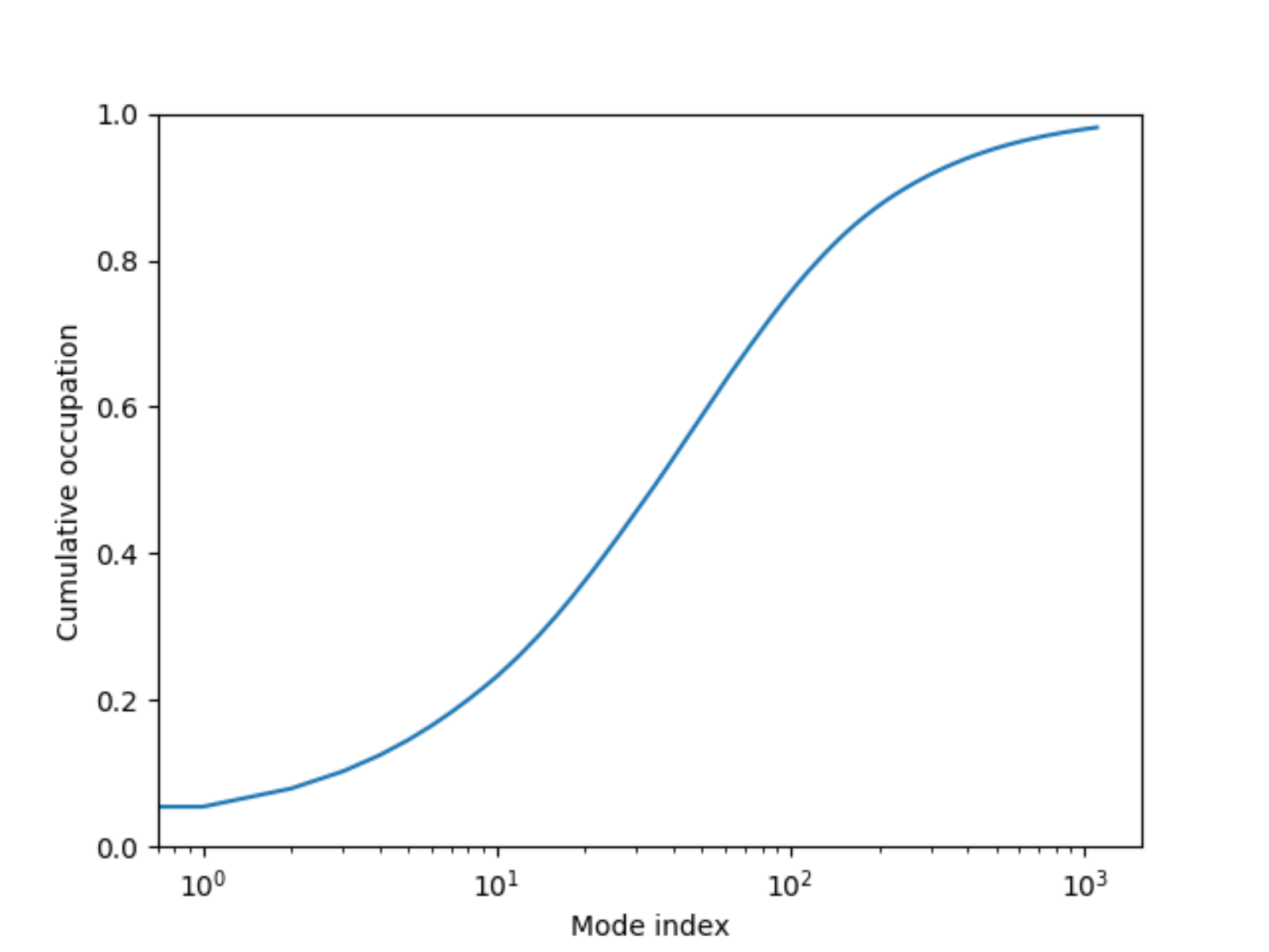}
\caption{Cumulative mode occupation for  emission of undulator U18 placed at EBS lattice and tuned to 17.226 keV. Occupation of lowest mode (coherent fraction) is 0.028.}
\label{cumulative_mode_occupation}
\end{figure}

The coherent mode decomposition for x-ray radiation emitted by a 1.4 meter long U18 undulator (period 18.3 mm), which is to be placed at the center of the straight section of the EBS (6 GeV, 147 pm~rad emittance storage ring), is here performed using COMSYL. The undulator is tuned to 17.226 keV ($K=0.411$) and the flux is 2.8$\times 10^{14}$ photons/s/0.1\%bw in a 1$\times$1 mm$^2$ aperture at 30 m. Figure~\ref{cumulative_mode_occupation} shows the cumulative mode occupation 
\begin{equation}
\begin{aligned}
\label{spectrum}
d_n(\omega)=\frac{\sum_{j=0}^{n-1} \lambda_j(\omega)}{\sum_{j'=0}^{\infty} \lambda_{j'}(\omega)}
\end{aligned}
\end{equation}
versus the total number of coherent modes $n$.

\begin{figure}
\includegraphics[width=9.0cm]{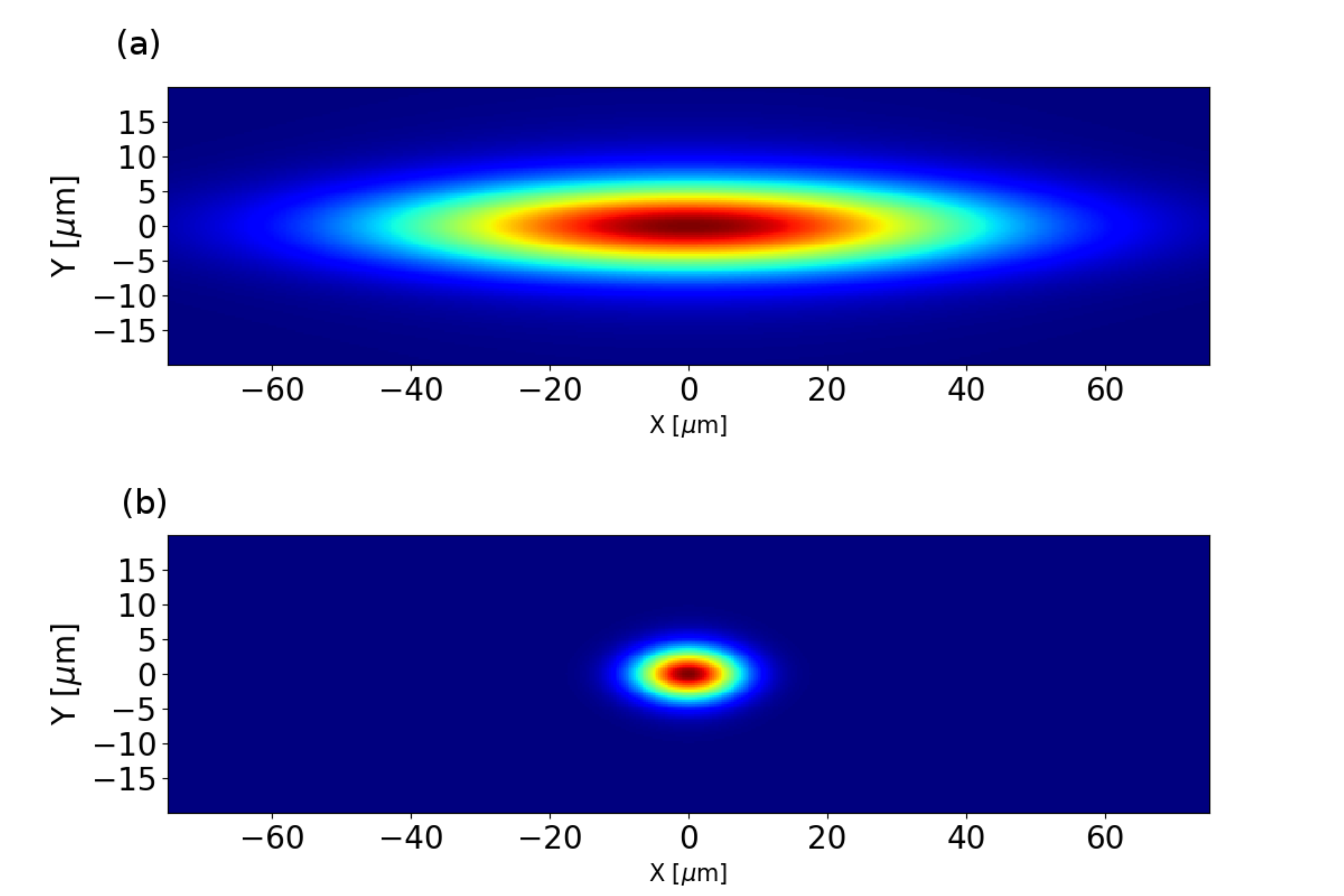}
\caption{(a) Spectral density (intensity) distribution at source plane. (b) Intensity of the first coherent mode at source plane.}
\label{spectral_density}
\end{figure}

The 1100 coherent modes calculated contain almost all (98\%) of the emitted radiation. The coherence fraction (occupation of first mode) is 2.8\%. The accumulation of the first 10 modes contains 21.7\% of the emitted intensity, followed by 33.0\% (20 modes), 73.3\% (100 modes) and 97.9\% (1000 modes). 
Figure~\ref{spectral_density} shows the spatial extension of the spectral density at the source plane, and also the extension of the first mode.

Figure~\ref{cumulative_mode_occupation} shows hundreds of modes are needed to represent more than 90\% of the spectral density. The spectral density FWHM (full width at half maximum) has dimensions 71.3$\times$10.9 $\mu$m$^2$. This agrees well with simple estimates considering the source size as a convolution of the undulator emission size ($\sigma_\gamma\approx (2.74/4\pi) \sqrt{\lambda L}\approx$9.6 $\mu$m, where  $\lambda$ is the radiation wavelength and $L$ is the undulator length) with the electron-bunch size ($\sigma_x$=30.2 $\mu$m, and $\sigma_y$=1.37 $\mu$m for the EBS straight section). The resulting FWHM values are 71.3$\times$10.0 $\mu$m$^2$. The first mode FWHM is 12.4$\times$6.11 $\mu$m$^2$.  

Note from Fig.~\ref{spectral_density} that the first coherent mode is non-zero (or, more precisely, non-negligible) over an area which is smaller than the area over which the spectral density is non-zero (non-negligible). Such a property will typically be true for a large class of partially-coherent beams, which will necessarily have at least two coherent modes since they are not fully coherent, by assumption.  This typical property follows directly from the fact that the spectral density of the beam is given by the sum of the spectral densities of each of the coherent modes, hence the area over which the spectral density is non-negligible is always non-contracting as we add progressively more coherent modes, irrespective of the particular non-zero eigenvalues associated with the modes.  

For the example in Fig.~\ref{spectral_density}a, the region over which the total spectral density is greater than some threshold value (say, 1\% of the maximum value) is roughly elliptical.  Denote this essential support of the spectral density by $\mathcal{S}$.  If the cross-spectral density were to vanish for any pair of points $(\vec{r}_1,\vec{r}_2)$, with $\vec{r}_1 \in \mathcal{S}$ and $\vec{r}_2 \in \mathcal{S}$, then this must be associated with a zero of the spectral degree of coherence $\mu$ (normalized cross-spectral density, see e.g.~p.~171 of the text by Mandel and Wolf \cite{mandel_wolf}) \cite{GburSPIE}. If there is one dominant coherent mode, it will typically be difficult to attain such zeros of the spectral degree of coherence, when both $\vec{r}_1$ and $\vec{r}_2$ lie within the region where the intensity of the first coherent mode is large.  This restriction vanishes when either or both of $\vec{r}_1$ and $\vec{r}_2$ lie outside the essential support of the dominant coherent mode.  

\begin{figure}
\includegraphics[trim=0 8mm 0 0,clip, width=7.5cm]{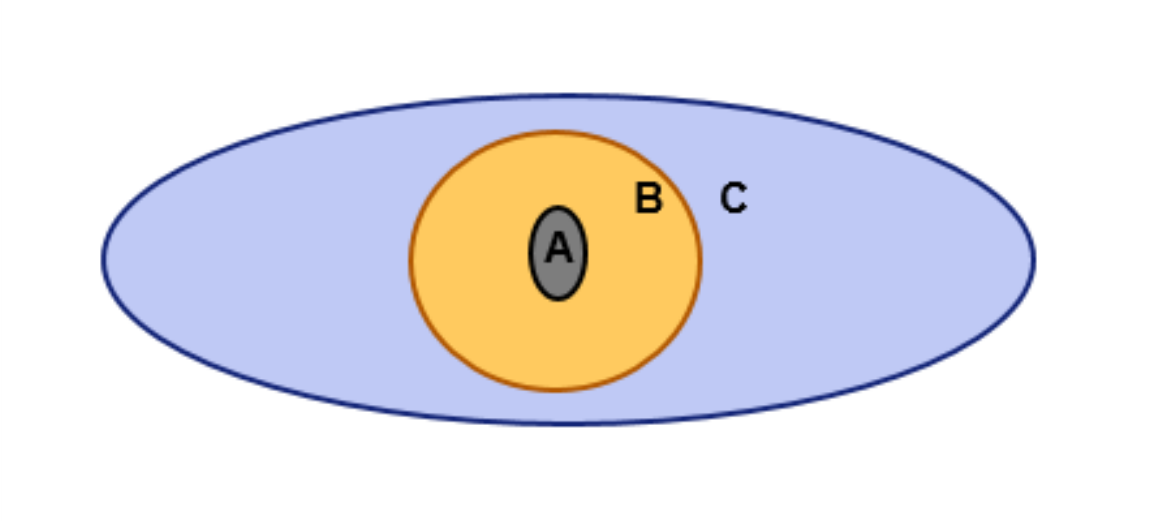}
\caption{Schematic representation of the different positions $(A,B,C)$ for the point $\vec{r_2}$, which is kept fixed in several subsequent figures (see text). The largest oval represents the spectral density (Fig.~\ref{spectral_density}a), and the inner circle the region over which the squared modulus of the first mode is non-negligible (Fig.~\ref{spectral_density}b). The gray oval represents the emission of a single electron. 
The points' coordinates (in $\mu$m) are: $A=(0,0)$, $B=(9.52,4.76)$ and $C=(20.83,9.82)$.}
\label{eye}
\end{figure}

Returning to the numerical simulation, our next task is to extract the CSD phase $\Phi$. Being a four-dimensional function when both $z$ and $\omega$ are fixed, it is convenient to fix two further variables (e.g.~components of the position $\vec{r_2}$) and observe how $\Phi$ depends on the other two transverse spatial coordinates. Several fixed points have been chosen: see Fig.~\ref{eye}. The first position (point $A$) is at the center of the emission. The second (point $B$) is at a non-centered position where the first mode has appreciable intensity. The third (point $C$) is outside the first mode, but in a place with appreciable intensity.

\begin{figure*}
\includegraphics[width=0.85\textwidth, trim=0 7mm 0 20mm,clip]{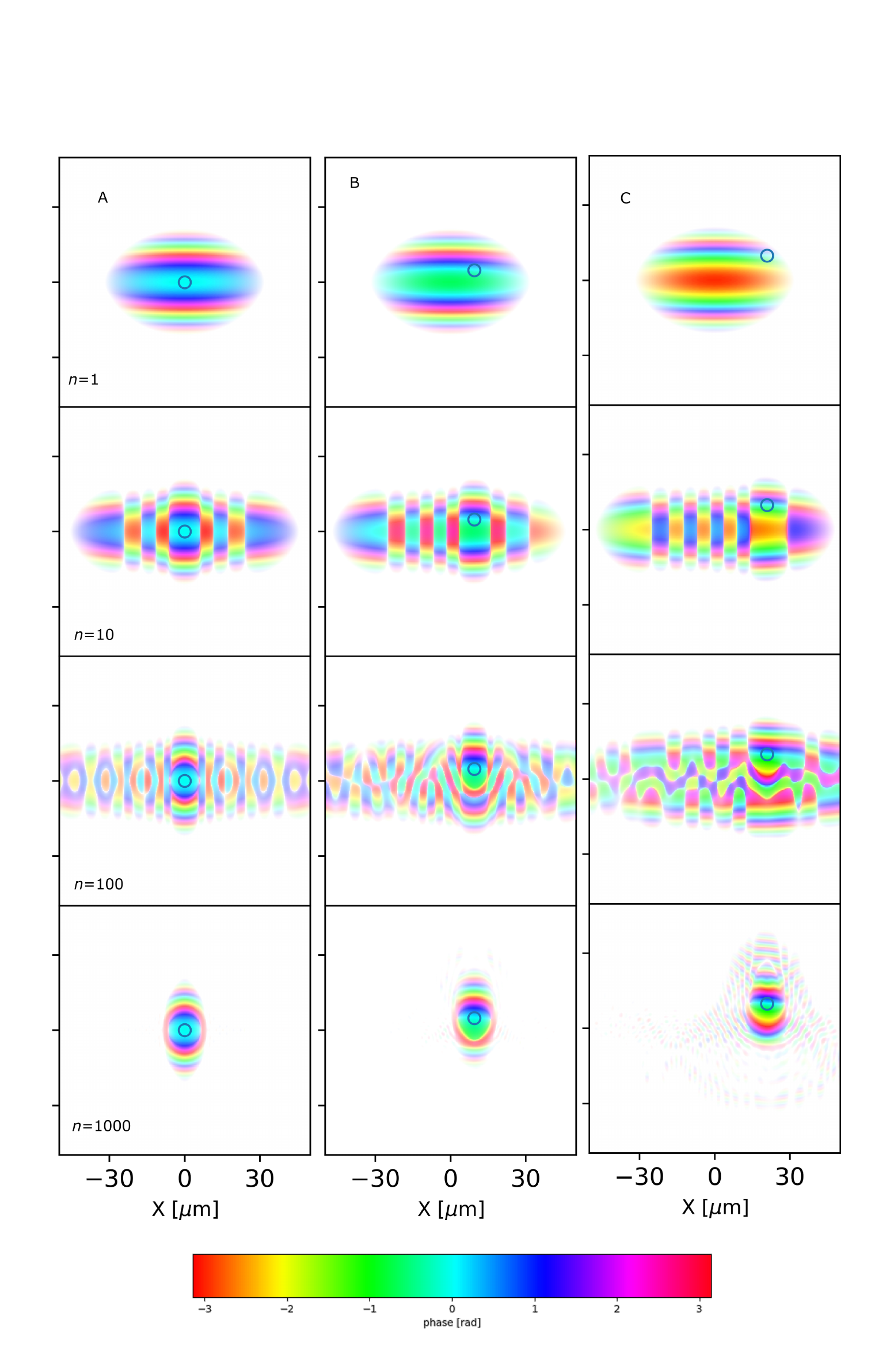}
\caption{Phase of $W(x,y,x_P,y_P)$ as a function of $(x,y)$ with fixed point $P=(x_P,y_P)=A$ (left column), $P=B$ (middle column) and $P=C$ (right column), when the number $n$ of coherent modes is (from top to bottom) 1, 10, 100, and 1000. The brightness of the displayed phase is proportional to $|W(x,y,x_P,y_P)|$, as CSD phase is not meaningful when $|W(x,y,x_P,y_P)|$ is negligible. In each image, a circle marks the position of the point $P$.}
\label{pointP}
\end{figure*}

When $\vec{r_2}$ is set to position $A$, we have the plot of the phase $\Phi(x,y,x_A,y_A)$ of the CSD in the left column of Fig.~\ref{pointP}, as a function of the coordinates $(x,y)$ of $\vec{r_1}$, with $z_1=z_2=z$ and $\hbar\omega$ all having the previously-indicated fixed values.  The brightness of the displayed phase has been taken to be proportional to $|W(x,y,x_A,y_A)|$, since CSD phase is not meaningful when $|W(x,y,x_A,y_A)|$ is negligible.  The figure shows four maps for the CSD phase, corresponding to the number $n$ of coherent modes being 1, 10, 100, and 1000 (top to bottom).  When only one coherent mode is included, there are no topological defects in the CSD phase.  However, when 10 coherent modes are included, domain-wall defects in the CSD appear.  Across each of these CSD-phase domain walls, the phase $\Phi$ of the CSD jumps by $\pi$ radians, with the CSD itself vanishing along each of the lines through which $\Phi$ changes discontinuously.  Adding more coherent modes, with $n=100$, the domain walls become curved and the region over which the CSD is non-negligible widens.  Finally, for $n=1000$, the effect of the increasing number of coherent modes is to narrow this region over which the CSD is non-negligible.  The general trend from top to bottom panels is a reduction in the spatial extent over which two-point field correlations have a magnitude $|W(x,y,x_A,y_A)|$ that is non-negligible, when one of the points is taken to be $A=(x_A,y_A)$, at the beam center.      

When $\vec{r_2}$ is set to position $B$, we have the plot of the CSD phase $\Phi(x,y,x_B,y_B)$ in the middle column of Fig.~\ref{pointP}, as a function of the coordinates $(x,y)$ of $\vec{r_1}$.  This figure again shows four CSD phase maps, corresponding to $n=1,10,100,1000$ coherent modes.  A similar trend to the left column is observed with regard to CSD domain walls.  However, in the middle column the additional feature of coherence vortices is present.  As we add more coherent modes, to the singularity-free case of $n=1$ in row 1 ($n=1$ mode), the topologically-unstable domain walls in row 2 ($n=10$ modes) begin to dissolve ($n=100$ modes) into  topologically stable CSD-phase defects, namely coherence vortices. As mentioned previously, an indicator of coherence-vortex cores is any points where all CSD phase-value colors converge like spokes on a wheel, with an associated vanishing of $|W(x,y,x_B,y_B)|$.  Such coherence vortices are evident in both the third ($n=100$ modes) and fourth ($n=1000$ modes) rows of the middle column.  For example, in the $n=1000$ case, the point $B$ (dark circle in the bottom row of the middle column) lies midway between a coherence vortex--anti-vortex dipole with topological charges of $m = \pm 1$. Another feature evident in the CSD is its speckled  structure in the $n=100$ case.  Such a ``patchy'' structure, which is also observed e.g.~in the Wigner function associated with chaotic quantum systems \cite{Zurek}, will influence quantities that are derived from the cross-spectral density via suitable coarse graining.

As a third and final example, when $\vec{r_2}=C$ we have the CSD phase maps $\Phi(x,y,x_C,y_C)$ in the right column of Fig.~\ref{pointP}.  Unlike the previous two examples, now the fixed spatial coordinate $C$ lies outside the dominant first mode's intensity distribution, but within the region where the spectral density of the entire beam is non-negligible.  We again see the previously-described trends, but with coherence vortices being somewhat more prevalent in other columns of the figure.  Also, in the $n=1000$ case, a weak speckled halo persists about the core patch where the CSD has non-negligible modulus.    

\section{Simulations of propagated x-ray undulator beams}

\begin{figure}
\includegraphics[width=9.0cm]{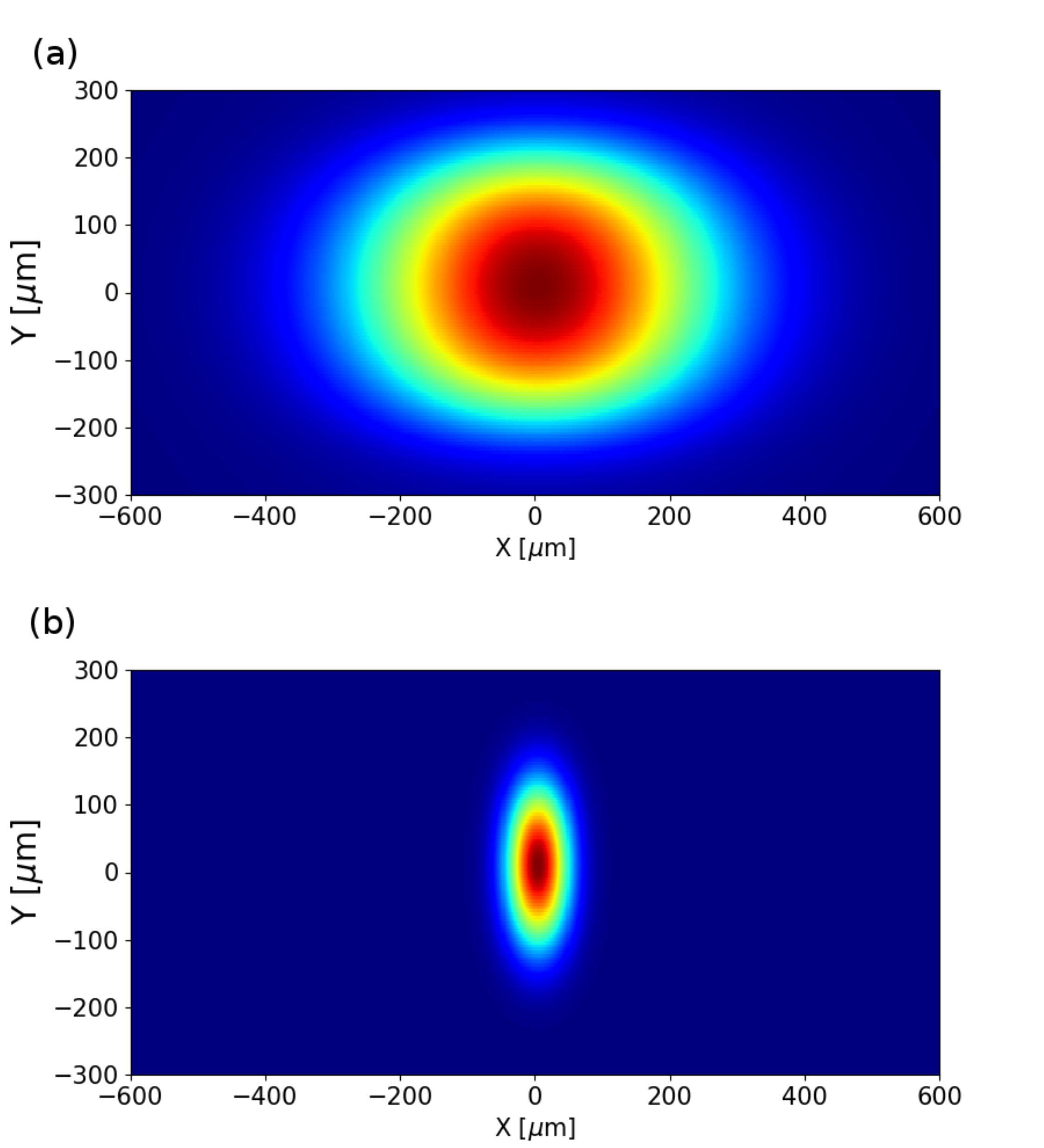}
\caption{(a) Spectral density (intensity) distribution over a plane placed 30 m from the source. (b) Intensity of the first coherent mode over the same plane.}
\label{spectral_density_propagated}
\end{figure}

In the previous section, calculation of the coherent mode decomposition for undulator radiation was performed in the source plane, located in the middle of the undulator. It is more realistic to propagate the radiation downstream from the source, e.g.~to a position in which a potential two-slit Young experiment would be feasible. As mentioned earlier, the propagated CSD is calculated by adding  propagated coherent modes. Propagation is performed using a Fourier representation of the Fresnel propagator, including a zoom factor \cite{schmidt,pirro} that enables adaptation of the window for different propagated distances. The propagated first mode and total spectral density at a distance $z=30$~m is shown in Fig.~\ref{spectral_density_propagated}. The FWHM of the spectral density is 431$\times$316 $\mu$m$^2$. Again, for a consistency check, this size can be compared with the propagation (30~m) of the beam divergence estimated as the convolution of the undulator emission divergence $\sigma'_\gamma=0.68\sqrt{\lambda/L}$ with the electron divergences ($\sigma'_x$=3.64 $\mu$rad and $\sigma'_y$=1.37 $\mu$rad giving 498$\times$366 $\mu$m$^2$ FWHM). 

Before proceeding further, recall that Fresnel diffraction of a coherent field through a distance $z>0$ results in development of a ``curvature of field'' coherent background.  This  gives a multiplicative factor of $\exp[ik(x^2+y^2)/(2z)]$ for the propagated field, where $k=2\pi/\lambda$: see e.g.~p.~16 of the text by Paganin \cite{paganin_book}.  When combined with the expression for the CSD in Eq.~\ref{eq:CSD}, we see that the CSD develops a corresponding multiplicative term $\exp[i\tilde{\Phi}]$ given by
\begin{eqnarray}
\nonumber\exp[i\tilde{\Phi}(x_1,y_1,z_1,x_2,y_2,z_2,\omega)] \quad\quad\quad\quad\quad\quad \\ =\exp\left[\frac{ik}{2}\left(\frac{x_2^2+y_2^2}{z_2}-\frac{x_1^2+y_1^2}{z_1}\right)\right].    
\label{eq:CSD-curvature-of-field}
\end{eqnarray}
The curvature of field $\tilde{\Phi}$ has been subtracted from all further CSD phase maps presented here, to avoid CSD domain walls and coherence vortices being rendered unclear by a strong continuous parabolic background, with respect to which such topological defects are invariant.  

\begin{figure*}
\includegraphics[width=0.85\textwidth]{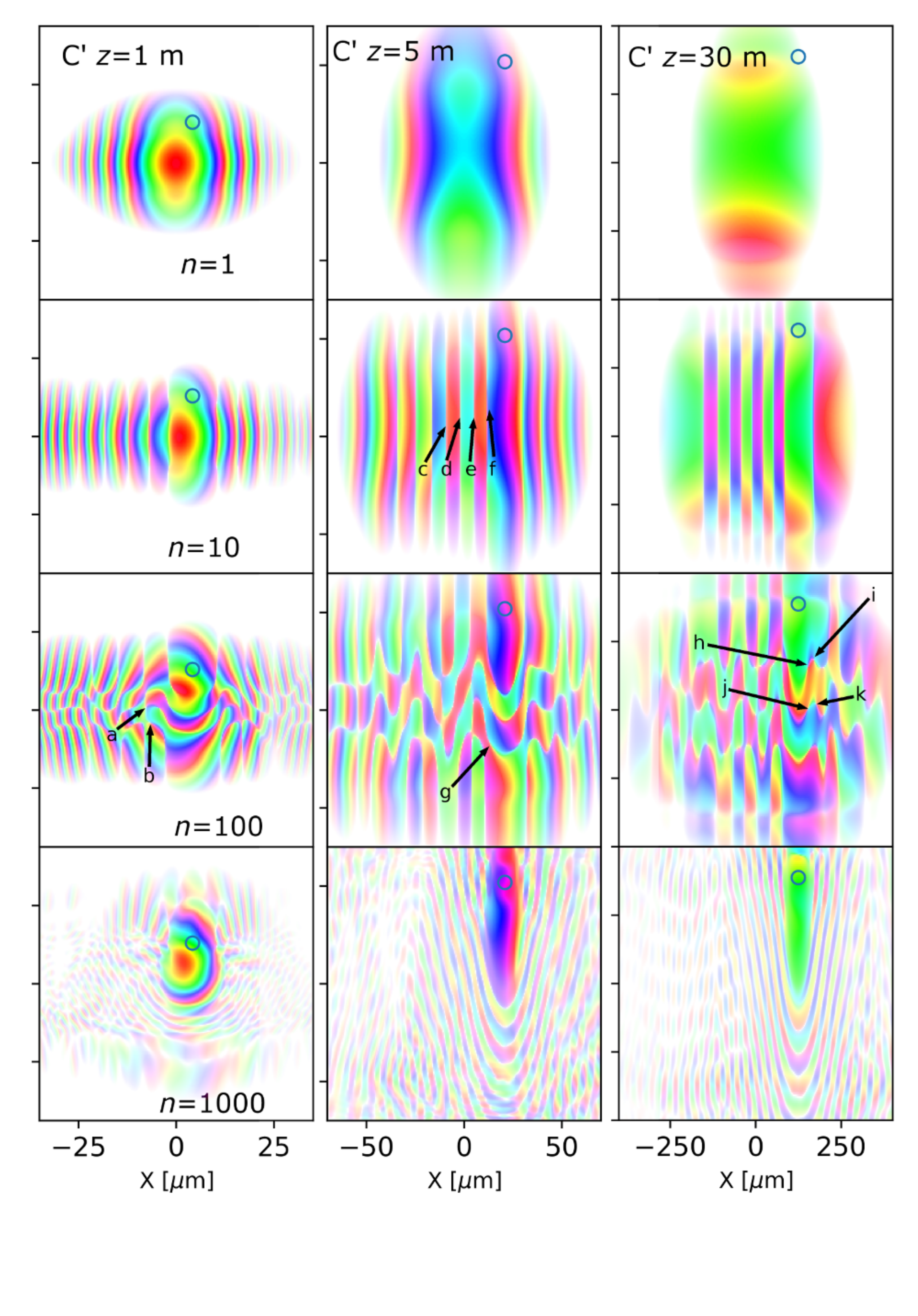}
\caption{Phase of $W(x,y,x_{C'},y_{C'})$ as a function of $(x,y)$ at $z$=1~m (left column), 5~m (middle column) and 30~m (right column), when the number $n$ of coherent modes is (a, top) 1; (b) 10; (c) 100; (d, bottom) 1000. The brightness of the displayed phase is proportional to $|W(x,y,x_{C'},y_{C'})|$, as CSD phase is not meaningful when $|W(x,y,x_{C'},y_{C'})|$ is negligible. In each image, a circle marks the position of the point $C'$. The curvature of field, as given in Eq.~\ref{eq:CSD-curvature-of-field}, has been removed from all plots.  The color table is as given in Fig.~\ref{pointP}.}
\label{pointC_propagated}
\end{figure*}

For the simulations in Fig.~\ref{pointC_propagated}, the CSD was propagated from the undulator source to distances $z=1,5,30$~m. The CSD phase $\Phi$ was calculated as a function of $\vec{r}_1$ for different fixed points $\vec{r}_2=A',B',C'$. These fixed points depend on $z$. For $z=30$~m, the coordinates have been selected to be homothetic with respect to the positions of $A, B, C$ at the source. That is, the FWHM$_x$ and FWHM$_y$ values at the source for the $x$ and $y$ directions, respectively, transversely scale linearly into the propagated ${\rm FWHM}'_x$ and ${\rm FWHM}'_y$ with ratio $\gamma_{x,y}(z=30)={\rm FWHM}'_{x,y}/{\rm FWHM}_{x,y}$.
For other distances $\gamma_{x,y}(z)= \gamma_{x,y}(z=30) z / 30.0$. Figure~\ref{pointC_propagated} shows the resulting CSD phase maps, for fixed point $C'$, with propagation distance $z$ increasing from left to right and the number of coherent modes $n$ increasing from top to bottom. These CSD phase maps are rich in both phase domain walls and phase coherence vortices.  (a) For $z=1$m (left column of Fig.~\ref{pointC_propagated}) the case of $n=1$ coherent modes is topologically trivial, but several almost-parallel domain walls form as soon as we pass to $n=10$ coherent modes.  These aligned domain walls partially dissolve as we move to $n=100$, with several coherence vortices forming, such as those marked $a$ and $b$.  Finally, for $n=1000$ coherent modes, the region of large $|W|$ in the vicinity of $C'$ has contracted to a smaller area, although there is a large halo of CSD speckle containing a significant fraction of the total area over which $|W|$ is non-negligible. (b) Similar trends are seen for the propagation distance $z=5$m (middle column of Fig.~\ref{pointC_propagated}): see e.g.~the parallel coherence domain walls marked $c,d,e,f$ and the coherence vortex marked $g$.  It is also interesting to note that, over the region where $|W|$ is non-negligible for $n=1000$ modes, the CSD phase becomes progressively flatter.  (c) Finally, we have $z=30$~m (right column of Fig.~\ref{pointC_propagated}), where the far-field regime is attained. In this regime, the shape of the magnitude and phase of the CSD is unchanged upon propagation, beyond a simple expansion with increasing $z$. A  vortex--anti-vortex dipole has been indicated by $hi$, with another such dipole at $jk$. 

\begin{figure*}
\includegraphics[width=0.8\textwidth]{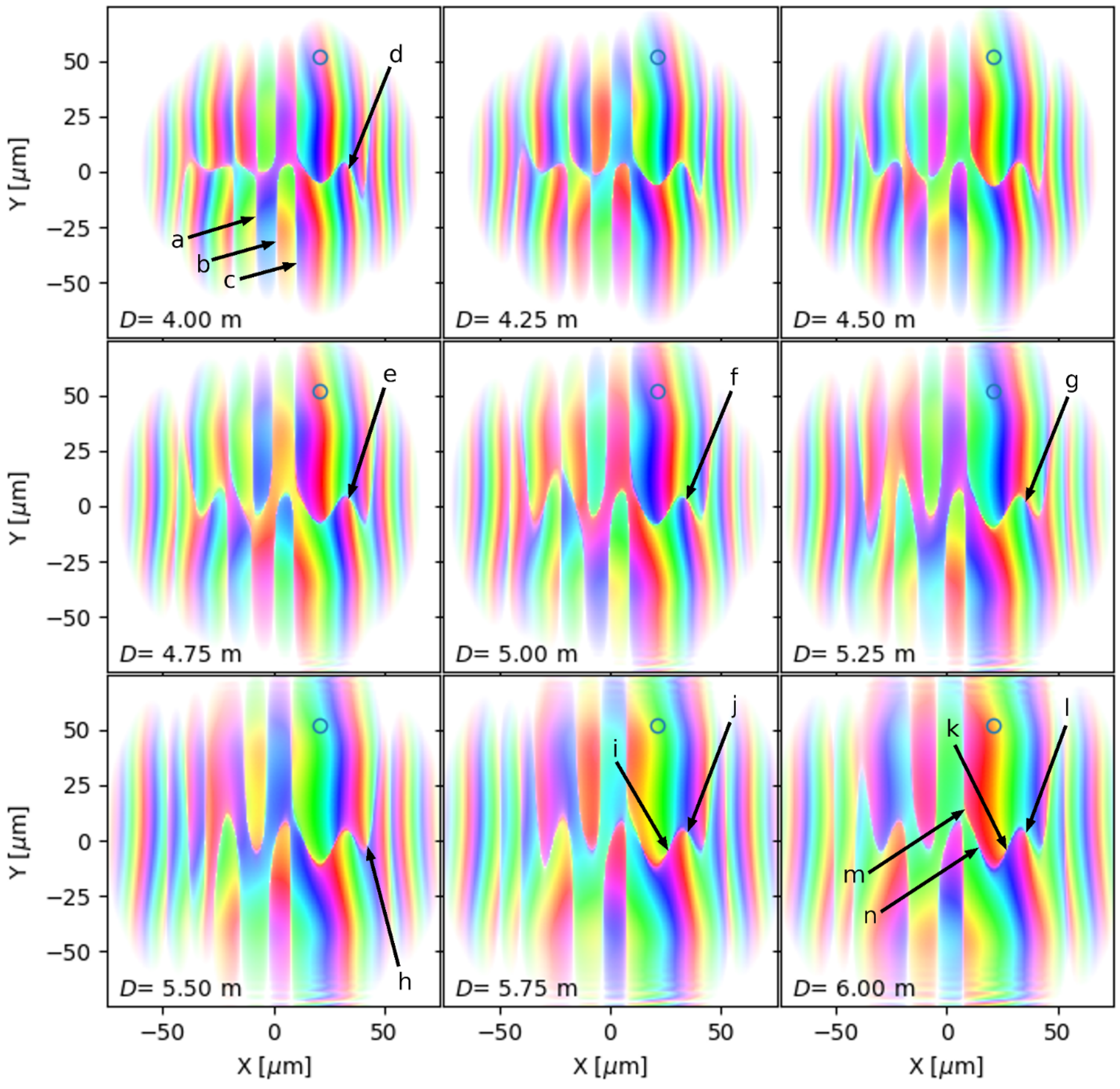}
\caption{Phase of $W(x,y,x_{C'},y_{C'})$ as a function of $(x,y)$ at different distances in the interval $z=5\pm 1$ m, with $n=20$ coherent modes. The brightness of the displayed phase is proportional to $|W(x,y,x_{C'},y_{C'})|$, as CSD phase is not meaningful when $|W(x,y,x_{C'},y_{C'})|$ is negligible. In each image, a circle marks the position of the point $C'$. The curvature of field, as given in Eq.~\ref{eq:CSD-curvature-of-field}, has been removed from all plots.  The color table is as given in Fig.~\ref{pointP}.}
\label{neighbour}
\end{figure*}

Next, we investigate how CSD phase maps evolve under free-space propagation \cite{TopologicalReactionsCohVortices,Marasinghe2010,Marasinghe2011}. See Fig.~\ref{neighbour}, which shows the propagated CSD phase $\Phi(x,y,x_{C'},y_{C'},z=D)$ for $n=20$ modes, with propagation distance $z=D$ varying from $D=4.00$~m to $D=6.00$~m in $0.25$~m steps, plotted as a function of $(x,y)$ with the point $C'=(x_{C'},y_{C'})$ kept fixed. This fixed point is indicated by a small circle, in all phase plots. The topologically-irrelevant curvature of field in the CSD phase, as given in Eq.~\ref{eq:CSD-curvature-of-field}, has again been removed from all plots. Coherence domain walls such as $a,b,c$ persist for all propagation distances, albeit with some distortion as $D$ increases.  Note the persistence of certain coherence vortices, but with some transverse displacement; see e.g.~the coherence vortex labelled $d,e,f,g,h$ which persists from frame to frame and which is therefore threaded by a core (CSD nodal line) like that in Fig.~2 of Marasinghe et al.~\cite{Marasinghe2010}. Such persistence is a direct consequence of the topological stability of CSD vortices.  A CSD coherence vortex--anti-vortex pair $ij$ is nucleated in passing from $D=5.50$~m to $D=5.75$~m, indicating a hairpin-shaped CSD nodal line \cite{Marasinghe2010,Marasinghe2011}.   This CSD dipole persists: see $kl$ at $D=6.00$~m.  In this last frame, another CSD vortex dipole has nucleated, labeled $mn$; a similar structure has been illustrated in a different context, in Fig.~7 of an earlier paper \cite{Marasinghe2010}.  Such CSD nodal-line dynamics, which are constrained by the topological conservation laws discussed earlier, evidence the dynamics permitted for such ``threads of incoherence'' \cite{TopologicalReactionsCohVortices, Marasinghe2011}.

\begin{figure}
\includegraphics[width=0.43\textwidth]{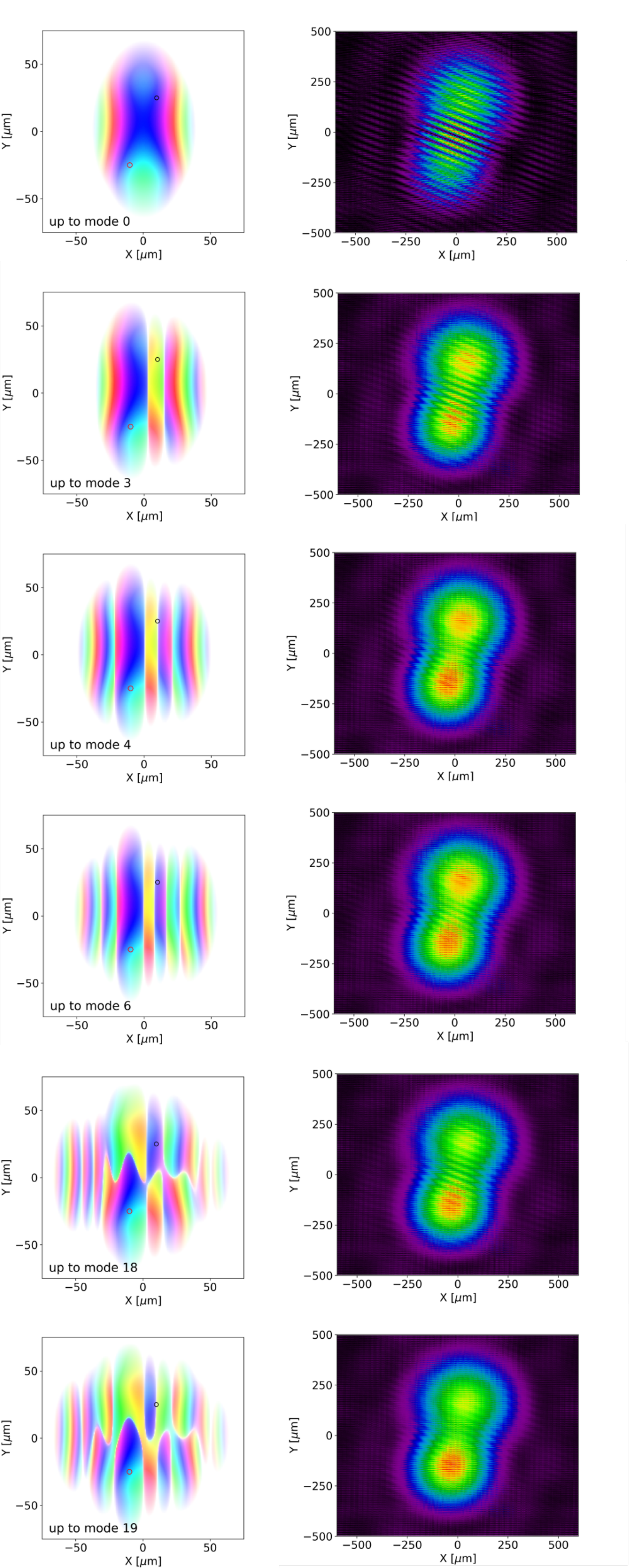}
\caption{On the left, the phase of $W(x,y,x_{D},y_{D})$ is shown. Circles represent the position of apertures at D (red) and E (black) used for a simulated Young experiment. On the right are interference patterns (more precisely, the spectral density obtained from the CSD) produced by propagating the radiation 30 m downstream from the plane containing the apertures. From top to bottom, the number of coherent modes used to build the CSD is varied (1, 4, 5, 7, 19, 20 modes respectively). The curvature of field, as given in Eq.~\ref{eq:CSD-curvature-of-field}, has been removed from all plots of CSD phase. The color table for all phase plots is as given in Fig.~\ref{pointP}.}
\label{young}
\end{figure}

Last, we  simulate a Young-type experiment by modelling the placement of two circular apertures in the plane to which the CSD has been propagated ($z$=5~m downstream of the source). These apertures are at points $(x_D,y_D)$=(-10~$\mu$m, -25~$\mu$m) and  $(x_E,y_E)$=(10~$\mu$m, 25~$\mu$m). The diameters of the apertures are 3.6 and 3~$\mu$m for points $D$ and $E$, respectively. The radiation passing through the apertures is numerically propagated 30~m further downstream. At the image plane a screen records the resulting spectral density (diagonal of the cross-spectral density). When the apertures are illuminated by a coherent source containing only the zeroth coherent mode, an interference pattern with high-visibility fringes is produced: see the top row of Fig.~\ref{young}. Note that the circles, in the phase plots of the CSD, represent the positions where the two pinholes are located.  For the single-mode case, there are no topological defects in the CSD phase.  Adding coherent modes in the construction of the CSD progressively reduces the visibility---see e.g.~the case of 4 coherent modes in row 2 of Fig.~\ref{young}.  In this second row we see a domain wall in the CSD phase, although this domain wall does not coincide with either pinhole.  When we move to the case of 5 coherent modes in row 3 of Fig.~\ref{young}, we notice that a {\em CSD domain wall now coincides with one of the pinholes}.  This heralds a dampening of the visibility in the associated Young-type fringes.  Such a dampening is evidenced by the fact that, as we add more coherent modes, the fringe visibility {\em improves} -- thus the fringe visibility with 19 modes (fifth row of Fig.~\ref{young}) is {\em higher} mid-way between the pinholes, than for the case of 5 modes in the third row.  This revival of fringe visibility corresponds to neither of the pinhole locations coinciding with a singularity in the CSD phase.  A CSD singularity again co-locates with one of the pinholes for the case of 20 coherent modes (bottom row of Fig.~\ref{young}), with a corresponding large drop in visibility when we pass from  19 coherent modes (second-bottom row of Fig.~\ref{young}) to 20 modes (bottom row of Fig.~\ref{young}).  Note also that there is a contrast reversal in the Young-type fringes, each time an additional CSD domain wall is interposed between the red and black circles in Fig.~\ref{young}.  For example, interferogram maxima are exchanged with minima, and vice versa, in passing from row 1 to row 2 of Fig.~\ref{young}.  

Although this procedure of adding modes to build the CSD is an artificial way to study fringe visibility depending on the beam coherence, in a real synchrotron experiment the beamline optics play a similar role. Some optical elements of the beamline (ideal reflectors, ideal focusing elements) do not alter the occupation spectrum of the modes. They are non-absorbing elements that conserve the Smith--Helmholtz invariant \cite{BornWolf}. However, if an optical element removes photons from the beam, or ``cuts'' intensity, then its effect is different for different modes. The new transformed ``modes'' can be used to build the CSD, but they cannot be considered coherent modes as they are no longer an orthonormal basis for expanding the CSD. A new coherent mode decomposition could be calculated on the transmitted CSD in order to obtain the new coherent modes. In the case of slits or pinholes centered on the optical axis, the lower-order coherent modes localized near the center of the beam axis will propagate, whereas the higher-order coherent modes that extend far from the axis will be absorbed. This will push the occupation spectrum to the lower-order modes with a consequent increase of coherence fraction but an obvious decrease of the spectral density (total intensity).     

\section{Discussion}

This discussion has three parts.  Section~\ref{subsec:Discussion-part-1} considers the influence of CSD correlation singularities on measured spectral densities.  Section~\ref{subsec:Discussion-part-2} explores the connection between unresolved speckle and the formalism of partially coherent light, with particular reference to the role of CSD correlation singularities.  Finally, Section~\ref{subsec:Discussion-part-3} outlines some possible avenues for future work.  
\subsection{Influence of CSD correlation singularities on measured spectral densities}\label{subsec:Discussion-part-1}

The speckled CSDs, considered in this paper, will influence measured intensity data in a variety of experiments.  This influence may be subtle.  We have already seen a specific example corresponding to a simple object composed of two point scatterers in Fig.~\ref{young}.  Here, we generalize to a broader class of scattering objects.

Consider the CSD scattered from a static deterministic non-magnetic sample with scattering potential \cite{Wolf1969,wolf_thin_book}
\begin{equation}
\begin{aligned}
\label{eq:ScatteringPotential}
F(x,y,z,\omega)=\frac{k^2}{4\pi}[n^2(x,y,z,\omega)-1],
\end{aligned}
\end{equation}
where $n(x,y,z,\omega)$ is the complex refractive index. The first Born approximation gives the scattered CSD, resulting from an incident CSD of $W^{(i)}$, as  \cite{wolf_thin_book}:
\begin{equation}
\begin{aligned}
W^{(s)} ({\bf r}_{\perp 1}, z_1=z, & {\bf r}_{\perp 2},z_2=z,\omega) \\  = \iiint_V \iiint_V &  W^{(i)}({\bf r}_{\perp 1}',z_1',{\bf r}_{\perp 2}',z_2',\omega)
\\ &\times F^*({\bf r}_{\perp 1}',z_1',\omega) \, F({\bf r}_{\perp 2}',z_2',\omega) \\ &\times G^*({\bf r}_{\perp 1},z_1=z;{\bf r}_{\perp 1}',z'_1;\omega) \\ &\times G({\bf r}_{\perp 2},z_2=z;{\bf r}_{\perp 2}',z'_2;\omega) \\ &\times d^2{\bf r}_{\perp 1}'~dz'_1~d^2{\bf r}_{\perp 2}'~dz'_2,
\label{eq:CSDScatteringFirstBornApproximation}
\end{aligned}
\end{equation}
where $G$ is the outgoing free-space Green function
\begin{equation}
\begin{aligned}
G( & {\bf r}_{\perp 1},z_1; {\bf r}_{\perp 2},z_2;\omega) \\ &=\frac{\exp[i \omega c^{-1}\sqrt{ | {\bf r}_{\perp 2}-{\bf r}_{\perp 1}|^2+(z_2-z_1)^2}]}{\sqrt{| {\bf r}_{\perp 2}-{\bf r}_{\perp 1}|^2+(z_2-z_1)^2}},
\end{aligned}
\end{equation}
$c$ is the speed of light in vacuum, ${\bf r}_{\perp j}\equiv(x_j,y_j)$ where $j=1,2$, $z=0$ is the exit surface of the sample,  $z_1,z_2\ge 0$ and $V$ is the volume occupied by the sample.

Now, the first Born approximation is a single scattering approximation, implying that the field is either scattered from a single point within the object, or is not scattered at all.  Equation~\ref{eq:CSDScatteringFirstBornApproximation} shows that, under this same approximation, the scattered {\em two-point correlation function} (i.e.~the CSD) may be viewed as being {\em singly scattered from every pair of points within the scattering volume}.  Thus, letting $({\bf r}_{\perp j},z_j)\equiv{\bf r}_j, j=1,2$, we see from Eq.~\ref{eq:CSDScatteringFirstBornApproximation} that (i) the incident CSD $W^{(i)}({\bf r}'_1,{\bf r}'_2,\omega)$ at the pair of points $({\bf r}'_1,{\bf r}'_2)$ is multiplied by the scattering potential $F^*({\bf r}'_1,\omega)F({\bf r}'_2,\omega)$, with (ii) the resulting correlations scattered from this pair of points then being propagated to a pair of points $({\bf r}_1,{\bf r}_2)$ via the double Green function $G^*({\bf r}_1;{\bf r}'_1;\omega)G({\bf r}_2;{\bf r}'_2;\omega)$, and finally (iii) the resulting scattered correlations being summed over every pair of points within the scattering volume. 

We now show that pairs of points, within the scattering volume, which correspond to a correlation-singularity zero of $W$, scatter in a fundamentally different manner to those that do not coincide with a correlation singularity.  To see this, consider a sample that consists of a pair of point-like scatterers, whose positions happen to coincide with the locations $A=({\bf r}_{\perp 1}'',z'')$ and $B=({\bf r}_{\perp 2}'',z'')$ associated with a coherence-vortex core. This amounts to Fig.~\ref{loss_of_fringe_visibility}a, with the screen removed and the pinholes at $A$ and $B$ being replaced with small scatterers at the same locations.  The associated scattering potential is:
\begin{equation}
\begin{aligned}
F({\bf r}_{\perp},z,\omega) &= \mathcal{P}(\omega)\, \delta({\bf r}_{\perp}-{\bf r}_{\perp 1}'',z-z'') \\ &+ \mathcal{Q}(\omega) \,\delta({\bf r}_{\perp}-{\bf r}_{\perp 2}'',z-z''),
\label{eq:ScatteringPotentialTwoDots}
\end{aligned}
\end{equation}
where the scattering amplitudes $\mathcal{P}$ and $\mathcal{Q}$ are complex functions of energy $\hbar\omega$, and $\delta$ is a Dirac delta.   Since $W^{(i)}({\bf r}_{\perp 1}'',z'',{\bf r}_{\perp 2}'',z'',\omega)$ vanishes due to the coherence vortex, and the Hermitian character of the CSD \cite{mandel_wolf} implies that $W^{(i)}({\bf r}_{\perp 2}'',z'',{\bf r}_{\perp 1}'',z'',\omega)$ also vanishes, substituting Eq.~\ref{eq:ScatteringPotentialTwoDots} into Eq.~\ref{eq:CSDScatteringFirstBornApproximation} leads to:
\begin{equation}
\begin{aligned}
\label{eq:CSDsforTwoScattersAtCVCore}
W^{(s)} ({\bf r}_{\perp 1}, &z_1=z, {\bf r}_{\perp 2},z_2=z) \\ = ~ &W^{(s)}_A ({\bf r}_{\perp 1}, z_1=z,  {\bf r}_{\perp 2},z_2=z) \\ &+ W^{(s)}_B ({\bf r}_{\perp 1}, z_1=z, {\bf r}_{\perp 2},z_2=z).
\end{aligned}
\end{equation}
Here, $W^{(s)}_A ({\bf r}_{\perp 1}, z_1=z, {\bf r}_{\perp 2},z_2=z)$ is the scattered CSD that would have been obtained if the scatterer at $A$ were to be present but scatterer $B$ were to be removed, with $W^{(s)}_B ({\bf r}_{\perp 1}, z_1=z, {\bf r}_{\perp 2},z_2=z)$ being the scattered CSD that would have been obtained if scatterer  $B$ were to be present but scatterer  $A$ were to be removed.  Upon setting ${\bf r}_{\perp 1}={\bf r}_{\perp 2}$ to convert the CSD to spectral density, we see that the spectral densities scattered from $A$ and $B$ merely add incoherently.  The interference term that would otherwise be present in the spectral interference law \cite{mandel_wolf} is suppressed by the coherence vortex: cf.~Fig.~\ref{young}.  If the pair of scatterers at $A$ and $B$, the scattering from which produces weak interference fringes, were to be rigidly transversely displaced or rotated in the beam, such that the pair no longer coincides with a coherence vortex, the interference term would re-appear.       

\begin{figure}
\includegraphics[width=8.5cm]{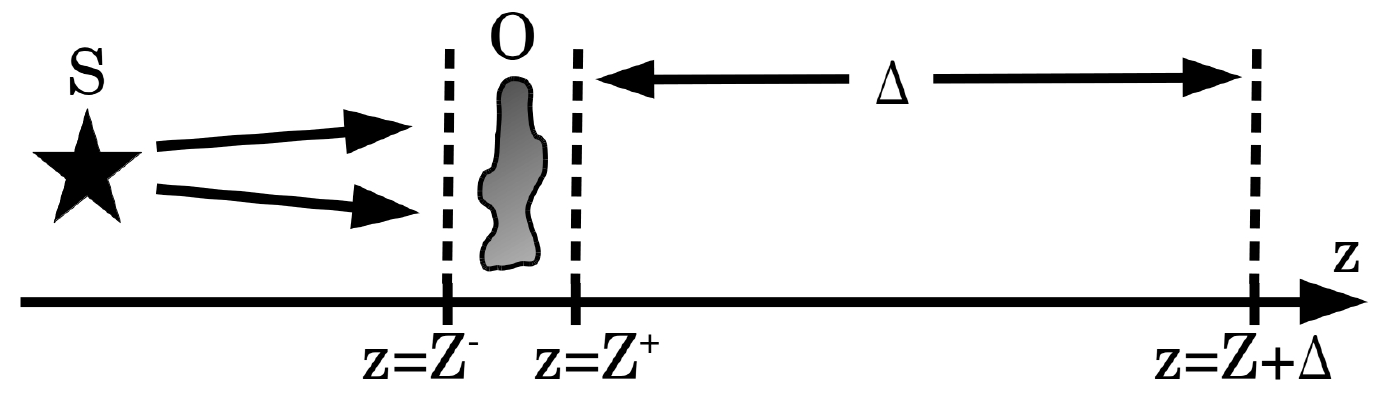}
\caption{A statistically stationary x-ray source $S$ generates a paraxial CSD illuminating the nominally planar entrance surface $z=Z^-$ of a thin object $O$.  The exit surface of the object is denoted by $z=Z^+$, with the CSD propagating from this exit surface through a free-space distance $\Delta$, to the surface $z=Z+\Delta$ of a position-sensitive detector.}
\label{fig:ImagingWithCoherenceVortex}
\end{figure}

For a more realistic example of the influence that coherence vortices have on measured x-ray spectral densities, consider {Fig.~\ref{fig:ImagingWithCoherenceVortex}}.  Here the previously-considered pair of point scatterers has been replaced with an arbitrary compact scattering distribution (i.e.~an ``object''), and the first Born approximation has been supplanted by the projection approximation \cite{paganin_book}.  The thin object  with complex transmission function $T({\bf r}_{\perp},\omega)$ is assumed to lie immediately upstream of the plane $z=Z^-$, with  object-to-detector distance $\Delta > 0$.  Use the coherent-mode expansion (Eq.~\ref{eq:CoherentModeExpansion}) to give the CSD in the plane $z=Z^-$ immediately upstream of the object, then apply the projection approximation to multiply each (paraxial) coherent mode by the complex transmission function of the object.  Next, use the Rayleigh--Sommerfeld diffraction integrals of the first kind \cite{Rayleigh, Sommerfeld, mandel_wolf} to propagate each coherent mode from the nominally planar exit surface $z=Z^+$ of the object to the detector plane $z=Z+\Delta$.  This gives the CSD for pairs of points on the detector surface:     
\begin{equation}
\begin{aligned}
\label{eq:CSD_with_object_and_detector}
W ({\bf r}_{\perp 1}, {\bf r}_{\perp 2},z_1 &=z_2=Z+\Delta,\omega) \\ =\sum_j \lambda_m({\omega}) & \left\{ [\psi_j({\bf r}_{\perp 1},\omega)~T({\bf r}_{\perp 1},\omega)]\star_1 K({\bf r}_{\perp 1},\Delta,\omega)\right\}^* \\ \times & \left\{ [\psi_j({\bf r}_{\perp 2},\omega)~T({\bf r}_{\perp 2},\omega)]\star_2 K({\bf r}_{\perp 2},\Delta,\omega)\right\},
\end{aligned}
\end{equation}
where 
\begin{equation}
K({\bf r}_{\perp},z,\omega)=-\frac{1}{2\pi} \frac{\partial}{\partial z} G({\bf r}_{\perp},z;{\bf 0}_{\perp},0;\omega) 
\end{equation}
is the Rayleigh--Sommerfeld convolution kernel, ${\bf 0}_{\perp}\equiv (0,0)$ and $\star_{1,2}$ denotes convolution with respect to ${\bf r}_{\perp 1}$ and ${\bf r}_{\perp 2}$ respectively.  Maps of the corresponding spectral density, obtained by setting ${\bf r}_{\perp 1}={\bf r}_{\perp 2}\equiv{\bf r}_{\perp}$ in the CSD, may be viewed as interferograms or inline holograms that are generated by many pairs of point scatterers.   

Whether we consider a pair of point scatterers under the first Born approximation (Eqs.~\ref{eq:CSDScatteringFirstBornApproximation}-\ref{eq:CSDsforTwoScattersAtCVCore}), or a thin object under the projection approximation (Eq.~\ref{eq:CSD_with_object_and_detector}), a similar conclusion is reached regarding the influence of correlation singularities on measured spectral densities: coherence vortices, and coherence domain walls, influence measured spectral densities.  

\subsection{Role of unresolved speckle in partial coherence}\label{subsec:Discussion-part-2}

Unresolved speckle underpins many phenomena exhibited by partially coherent classical optical fields \cite{Nugent2003,paganin_book,Nesterets2008}.  Here, speckles are considered to be unresolved if they are coarse grained over space (via the spatial extent of a detector pixel) or coarse-grained over time (via the acquisition time during which photons are detected), to an extent that either (i) no speckles are detected or (ii) the length and/or temporal scales of the {\em measured} speckles exceed those of the underpinning speckle fields.  ``Speckle'' is here used to describe functions that exhibit rapid intensity variation with respect to spatial (and, where appropriate, temporal) variables. This definition is more suitable for our purposes than merely equating the term ``speckle'' with ``fully developed speckle''. 

As a first example, consider the time-dependent and position-dependent intensity $I(x,y,t)=|E(x,y,t)|^2$ of a partially coherent beam-like $z$-directed paraxial field illuminating a region $\Omega$ of a 2D detector in a plane of constant $z$.  At each instant of time, the field will typically be a highly speckled function of transverse position $(x,y)$.  These speckles will typically evolve appreciably over time-scales on the order of the coherence time.  The characteristic transverse extent of these speckles may be rather small, and---for the case of fully developed speckle, which is in fact {\em typical} when considering the {\em instantaneous} intensity $I(x,y,t)$ of a classical partially coherent scalar field---on average there will be about one vortex in the instantaneous phase $\arg  E(x,y,t=t_0)$, for each speckle.  These intensity speckles are coarse-grained in space and time, due to both the pixel size and acquisition time of the detector.  Phase-vortex velocities will typically evolve in a highly non-linear and indeed chaotic manner with time \cite{Alperin2019}. The {\em spatio-temporally coarse-grained intensity distribution is manifest as partial coherence}, namely a loss of the maximal visibility associated with fully developed speckle \cite{paganin_book}.  

The concept of unresolved speckle harmonises with the idea that {\em spatio-temporal coarse-graining influences the degree to which the coherence of the field affects measured intensity data.}  We saw an example of such coarse-graining of dynamic spatio-temporal speckles, in the previous paragraph.  Now consider the simpler case of static fully-developed coherent speckle. Such a speckled beam would typically be infused by a spatially random ``gas'' of phase vortices, with an equal number of clockwise and anti-clockwise vortices, one per speckle \cite{Goodman2007,paganin_book}.  Classically, the field intensity vanishes at vortex cores \cite{Dirac1931}, which may be viewed as  exhibiting a high degree of coherence since the visibility of the generalized interference ``fringes''---namely the intensity speckle---approaches unity.  However, this near-unity visibility will only be manifest {\em if one's detector has sufficiently fine pixels}, namely pixel dimensions significantly smaller than the transverse length scale associated with the speckles.  If, conversely, exactly the same temporally-static fully-developed coherent speckle field were to have its intensity measured with a pixellated detector whose spatial dimensions are much larger than the characteristic speckle size, the resulting coarse-grained intensity map would be smoother, hence of correspondingly lower visibility and lower ``coherence''.  What is perhaps surprising about correlation singularities, when viewed from this perspective, is their association with pairs of points for which there is not just low coherence, but {\em zero} coherence.     

For a third and final example, of the fact that spatio-temporal coarse-graining influences the effective degree of coherence that is manifest in optical experiments utilizing partially coherent radiation, recall the experiment of Magyar and Mandel \cite{MagyarMandel1963}.  This studies Young-type interference fringes produced by superposing independent maser beams.  Here, ensemble-averaged quantities such as the CSD are inapplicable.  For two independent quasi-monochromatic sources, whose relative phases drift over times on the order of the coherence time:~(i) we will measure Young-type fringes in a random position if the exposure time is shorter than the coherence time and there are enough photons registered to form an image; (ii) these fringes will be washed out if the exposure time is significantly longer than the coherence time.  Also, (iii) even if the exposure time is sufficiently small compared to the coherence time for fringes to persist after temporally averaging over the measurement interval, they will only be resolved if the spatial averaging implied by the pixel size does not smear the said fringes away.  This may be immediately applied to a correlation-singularity context: If two points in space are such that the magnitude of $W$ vanishes at some given angular frequency for that particular pair of points, then (i) the combination of disturbances scattered from each point will yield time-averaged Young-type interference fringes with suppressed visibility (as seen in Fig.~\ref{young}), but (ii) if the intensity were to be averaged over timescales much shorter than the coherence time, instantaneous Young-type fringes of stronger visibility would be observed.   

Since we are working with a space--frequency description of partially coherent radiation \cite{Wolf1982,mandel_wolf,wolf_thin_book}, the time variable has been Fourier transformed away.  Speckles, initially present in the physical fields underpinning the calculation of a given CSD, manifest as the speckled CSD structures that will often be present in the CSD for realistic sources such as the modern x-ray undulator considered here.  This nuances the concept of a coherence area, in a sense that we now describe.  The area of the region $\Omega$ in $(x,y)$ space, centered on $(x_0,y_0)$, where $|W(x_0,y_0,x,y)|$ is non-negligible, defines a coherence area in the usual sense of the term.  However, if the patch $\Omega$ of $xy$-space (where $|W(x_0,y_0,x,y)|$ is non-negligible) possesses speckled phase structure then it is partitioned into cells of a {\em second, smaller characteristic length scale associated with the CSD speckles}.  These CSD speckles are cells bounded by correlation singularities where the CSD vanishes, and across which the CSD phase changes markedly.  As a simple example of this, the second row of Fig.~\ref{pointP} has the ``first coherence area'' (region where $|W|$ is non-negligible) partitioned into 9 (left column) or 10 (middle and right columns) fragments that are separated by domain walls.  More complicated examples of the same idea are given in CSD phase maps such as those in the third row of Fig.~\ref{pointC_propagated}.

\begin{figure}
\includegraphics[width=8.5cm]{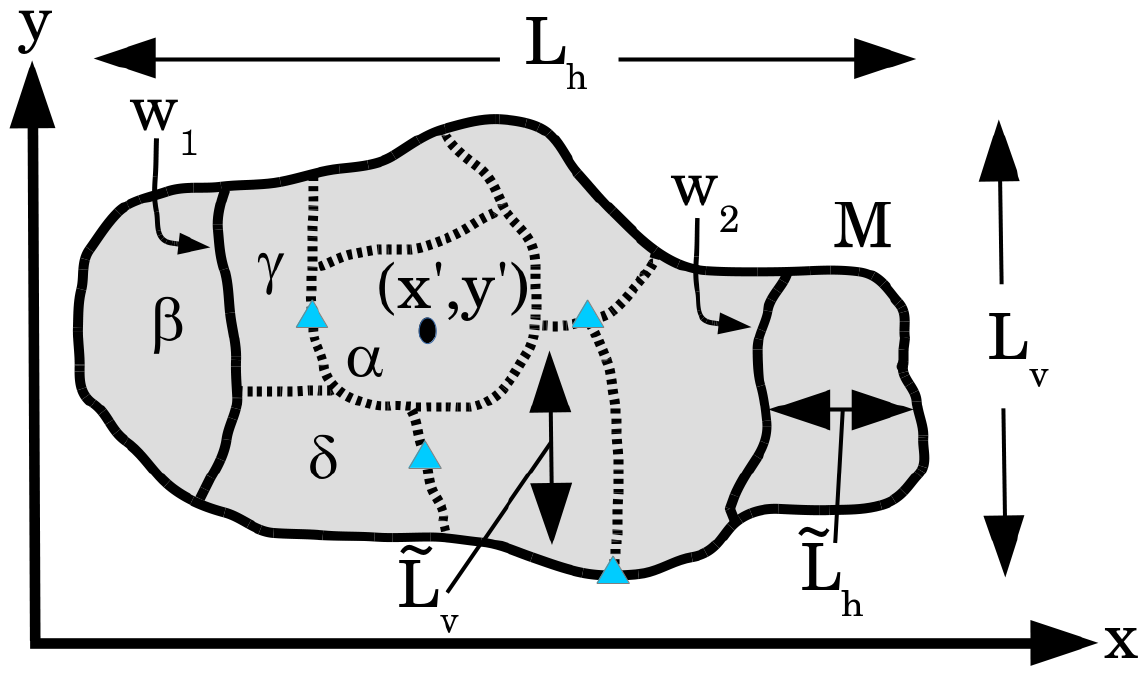}
\caption{Speckle structure associated with coherence area.  Shaded in gray is the manifold $M$ of points  over which $|W(x,y,x',y',z=z',\omega)|$ is non-negligible, for fixed $x',y',z',\omega$.  This coherence area has characteristic horizontal and vertical scales $L_{\textrm h}$ and $L_{\textrm v}$ respectively.  When speckled, the coherence area may be partitioned into cells $\alpha, \beta,\gamma,\delta,\cdots$, between which the CSD phase fluctuates significantly.  Such cells, of respective characteristic horizontal and vertical scales $\tilde{L}_{\textrm h},\tilde{L}_{\textrm v}$, may be flanked by CSD phase domain walls such as $w_1,w_2$ and/or coherence vortices shown as triangles.}
\label{CoherenceAreas}
\end{figure}

The ideas of the preceding paragraph are abstracted in Fig.~\ref{CoherenceAreas}.  Shaded in gray is the manifold of $(x,y)$ points $M$ at which the CSD $W(x,y,x',y',z=z',\omega)$ is non-negligible, with $(x',y')$ and energy $\hbar\omega$ being fixed, in a given plane of fixed $z=z'$.  The vertical transverse coherence length $L_{\textrm v}$ and horizontal coherence length $L_{\textrm h}$ are as indicated, corresponding to coherence area
\begin{equation}
    \mathcal{A}=L_{\textrm v}L_{\textrm h}.
\end{equation}
This coherence area is appropriate insofar as light from any point $(x,y)\in M$ will be able to exhibit interference fringes of non-zero visibility when combined with light from $(x',y')\in M$, at energy $\hbar\omega$, provided that $(x,y,x',y')\in M \times M \backslash \Upsilon$ does not coincide with CSD correlation singularities such as the domain walls $w_1$ and $w_2$ or the CSD vortices indicated by triangles. The ``coherence patch'' $M$ is broken into $\mathcal{N}$ coherence cells labeled $\alpha,\beta,\gamma,\delta$ etc., with the CSD phase varying significantly between such domains.  Hence the position of the interference fringes, resulting when light from two different domains (e.g.~$(x',y')\in\alpha$ and $(x,y)\in\beta$), will be  affected by the phase difference between domains.  Also, if light from many pairs of points straddling many different coherence cells is combined (e.g.~via Eqs.~\ref{eq:CSDScatteringFirstBornApproximation} or  \ref{eq:CSD_with_object_and_detector}), coherence effects may be suppressed by the many essentially random phase shifts corresponding to different cells.  These cells have characteristic vertical and horizontal transverse dimensions $\tilde{L}_{\textrm v},\tilde{L}_{\textrm h}$ respectively, corresponding to a second relevant coherence area
\begin{equation}
    \tilde{\mathcal{A}}=\tilde{L}_{\textrm v} \tilde{L}_{\textrm h}.
\end{equation}
The number of cells into which $M$ is partitioned, which may be viewed as an order parameter \cite{SethnaBook}, is:
\begin{equation}
    \mathcal{N}=\mathcal{A}/\mathcal{\tilde{A}}.
\end{equation}

As mentioned earlier, we may speak of coherence-vortex cores as coordinate pairs for which there is ``complete destructive interference of coherence''. This is because of the vanishing of $W$ for pairs of points that correspond to a correlation singularity.  The web of CSD zeroes  associated with coherence vortices is five-dimensional, and embedded in seven dimensions corresponding to the coordinates $(x,y,z,x',y',z',\omega)$ of $W$ \cite{Marasinghe2010}. CSD zeroes associated with domain walls are six-dimensional.  This web of incoherence, embedded within the CSD, is naturally formed rather than being an exotic construct, in a manner not unrelated to the spontaneous formation of a gas of phase vortices in the complex wave-field associated with fully developed speckle \cite{OHolleran2008}, or the spontaneous formation of phase vortices in the focal volume of an aberrated lens \cite{BoivinDowWolf1967}.  Such a web of incoherence influences, for example, spectral densities calculated by setting both spatial coordinates equal after the integration in Eq.~\ref{eq:CSDScatteringFirstBornApproximation}.  Moreover, an uncountable infinity of pairs of points, within a given scattering volume, generates scattered radiation that is incoherently superposed, even though the incident radiation is partially coherent.

\subsection{Further work on x-ray coherence vortices}\label{subsec:Discussion-part-3}

\subsubsection{Topological reactions of x-ray correlation singularities} 

An interesting topic for future work would be the {\em topological reactions} associated with x-ray coherence vortices and domain walls  \cite{GburSPIE,TopologicalReactionsCohVortices,Marasinghe2010}.  We already saw an example of this in Fig.~\ref{neighbour}.  Such topological reactions include the annihilation of a clockwise coherence vortex with an anti-clockwise coherence vortex and the spontaneous creation of a clockwise--anti-clockwise pair of coherence vortices \cite{TopologicalReactionsCohVortices}.  Other reactions are possible such as the decay of higher-order coherence vortices to multiple lower-order coherence vortices \cite{TopologicalReactionsCohVortices}, together with mutual annihilation of a CSD phase saddle-point and a local phase maximum or minimum.  Note that the topological conservation laws for coherence vortices (i.e.~conservation of topological charge) must be augmented with a second conservation law (conservation of topological index) when coherence vortices are considered in relation to maxima, minima and saddle points of CSD phase.  See e.g.~Mays et al.~\cite{MaysPonsaingPaganin2018}, and references therein, for further information on such additional conservation laws. 

\subsubsection{Tensorial x-ray correlation defects} 

Correlation singularities assume a more general character when the vector nature of the electromagnetic field cannot be ignored.  In such cases, the cross-spectral density generalizes from a complex scalar field to a tensor field \cite{mandel_wolf,wolf_thin_book}.  This correlation tensor is of second rank for paraxial fields, and third rank for non-paraxial fields.  Note also, that when considering partial coherence for vector electromagnetic fields, partial polarization should also be taken into account \cite{wolf_thin_book}.  We already saw that scalar cross-spectral densities admit topological defects such as CSD-phase coherence vortices and domain walls; similarly, CSD tensors associated with partially coherent (and partially polarized) vectorial electromagnetic fields also admit defects.  Permissible defects in correlation tensors are classifiable via homotopy theory \cite{VilenkinShellard1994,Volovik2003}.  Coherence skyrmions, coherence textures, coherence simplices \cite{SimulaPaganin2012} and other exotic tensorial defects are to be expected.

\subsubsection{Pulsed fields}

Periodically pulsed fields, rather than being considered statistically stationary, may instead be cyclo-stationary \cite{Gardner,Schoonover2009}.  The cross-spectral density is then a function of two angular frequencies, rather than one.  It would be interesting to consider correlation defects in this more general setting, e.g.~in the context of periodically pulsed sources such as x-ray Free Electron Lasers.  

\subsubsection{Experimental observation} 

To the best of our knowledge, to date there is no direct experimental evidence for CSD vortices and CSD domain walls in the x-ray domain.  Experiments to observe x-ray coherence vortices, both directly and indirectly, therefore warrant  attention. There could be several reasons why these singularities have not been detected so far. One reason is the difficulty of working with circular apertures with diameter smaller than the characteristic size of a correlation singularity core (for CSD vortices) or domain-wall width (for CSD domain walls). It would be desirable to position these small-diameter pinholes at a pair of positions where a correlation singularity exists. However, for practical reasons this is difficult, thus an aperture with finite size averages the CSD over an area that includes more than the CSD vortex or CSD domain wall. This is clear in our simulations ({Fig.~\ref{young}}) where there is an appreciable change in visibility when a singularity enters the aperture pair, but it is not as dramatic an effect as might be expected (see e.g., {Fig.~\ref{Young_fringe_anholonomy}}, curve 1). It may also be possible that the effect of singularities in the CSD has degraded the coherent properties of existing beams without being recognized as such in experiments. 

The implementation of upgraded storage-rings at x-ray synchrotron facilities such a EBS at ESRF, where the coherent fraction would improve roughly by two orders of magnitude, opens the possibility for new experiments. The partially-coherent flux through small pinholes will certainly improve, thus the CSD-vortex-sensitive Young experiment as discussed in this paper may become feasible. If a beamline images the source as a secondary source where a slit is placed, the slit aperture would act as a coherent-mode filter: the more closely spaced the aperture, the smaller the number of coherent modes that would be transmitted. Therefore, a further screen with a double pinhole and a detector could reproduce the situation presented in {Fig.~\ref{young}}. Moreover, if one pinhole can move in a trajectory around a singularity, ratcheting interferometers could be obtained, as sketched in Fig.~\ref{Young_fringe_anholonomy}. 

The experimental observation of x-ray correlation singularities has a more applied aspect beyond those considered above.  Rather than considering x-ray correlation singularities as objects that are at the focus of certain experimental studies---as has been done in both the preceding paragraphs (for the x-ray case) and also in previous works (for the visible-light \cite{Bogatyryova2003,Palacios2004,Wang2006,Ambrosini2005} and acoustic \cite{Basano2005} cases)---we can instead ask the following question. In view of the complexities involved in computation, can we ever expect to be able to disentangle the effects of correlation singularities from imaging or diffraction data, so as to reconstruct samples illuminated by partially coherent x-ray beams that contain such singularities?  A first step towards such disentanglement is to establish the role of these singularities in observable x-ray intensity data, which has been a main theme of the present paper. The effects are subtle but measurable, according to our realistic simulations where the parameter controlling the partial coherence is the number of coherent modes. This in turn implies that correlation singularities have an effect upon reconstructions that are obtained based on intensity data that are themselves influenced by correlation singularities. In real x-ray experiments, it is well known that a lack of complete beam coherence influences intensity images and makes it necessary to take into account several coherent modes, e.g.~in ptychographic reconstruction  ~\cite{Stockmar2015}. In this sense, x-ray correlation singularities are expected to play at least an implicit role in object-reconstruction procedures.  In this context,  ptychographic paradigms \cite{Rodenburg2008} have the strength of data redundancy: since each point on the entrance surface of the sample is illuminated more than once, information that may be missing from one image due to the presence of correlation singularities that suppress interference between particular pairs of points in a given illumination patch, may be ``filled in'' with information from an adjoining illumination patch.  It is an interesting open question to ask whether correlation singularities might be more challenging to account for, either implicitly or explicitly, in single-shot approaches to x-ray inverse imaging problems such as the inside-source method for x-ray Fourier holography \cite{FaigelTegze1999} or x-ray coherent diffractive imaging \cite{CDI}.  Such interesting avenues for future investigation give context to the present work, on account of the previously mentioned fact that a first step towards an ability to address the inverse problem---of disentangling the effects of correlation singularities upon an object that is reconstructed from measured intensity data obtained using partially coherent x-ray beams---is to have an accurate forward-problem model for how those intensity data are influenced by such x-ray correlation singularities that may be present in the illuminating beam.

\section{Conclusion}

Coherence vortices and domain walls will exist in many non-trivial x-ray fields.  Such correlation singularities influence the images that one takes, in ways that may lead to misleading results if one simply ignores their existence.  Correlation singularities were seen via simulation to be present in the field generated by a modern x-ray undulator. Such singularities, which are not present in many simple models for partially coherent sources, were seen in our model to imply a speckled structure in the associated cross-spectral density.  Coherence vortices were seen to persist even if the most populated coherent mode has a relatively large fraction of the total optical power, corresponding to a high coherent fraction and a source that has a high degree of coherence. In light of this investigation, the concept of a single transverse coherence length was extended.   Some avenues for future work were sketched.  

\section*{Acknowledgements}

We acknowledge useful discussions with Mario Beltran, David Ceddia, Carsten Detlefs, Mark Glass, Kieran Larkin, Kavan Modi, Kaye Morgan and Tim Petersen. DMP acknowledges financial support from the European Synchrotron Radiation Facility, and the University of Canterbury.  

\bibliography{iucr}

\end{document}